\documentclass[aps,prl,twocolumn,superscriptaddress]{revtex4-1}

\usepackage{amssymb,amsfonts,amsmath}

\usepackage{graphicx}
\usepackage{amssymb}
\usepackage{epstopdf}
\usepackage{amsmath}
\usepackage{amsfonts}   % if you want the fonts
\usepackage{amssymb}    % if you want extra symbols
\usepackage{graphicx}
\usepackage{longtable}
\usepackage{float}
\usepackage{color}

\newcommand{\G}{{\bf G}}

\newcommand{\smll}[1]{\mbox{\footnotesize $#1$}}

\begin{document}

%\title{Inferring bulk dynamics from boundary measurements}
\title{Inferring Epigenetic Dynamics from Kin Correlations}

\author{Sahand Hormoz}
\affiliation{Kavli Institute for Theoretical Physics, Kohn Hall, University of California, Santa Barbara, CA 93106, USA}
\author{Nicolas Desprat}
\affiliation{Laboratoire de Physique Statistique (UMR8550), Ecole Normale Superieure, 24 rue Lhomond, 75005 Paris, France}
\affiliation{Institut de Biologie de l'ENS (IBENS UMR 8197), Ecole Normale Superieure 46, rue d'Ulm 75230 Paris, France}
\affiliation{Universite Paris Diderot 75205 Paris CEDEX 13, France}
\author{Boris I. Shraiman}
\affiliation{Kavli Institute for Theoretical Physics, Kohn Hall, University of California, Santa Barbara, CA 93106, USA}
\affiliation{Department of Physics, University of California, Santa Barbara, CA, 93106, USA}

\renewcommand{\today}{February 3, 2015}
\date{February 3, 2015}

\begin{abstract}
Populations of isogenic embryonic stem cells or clonal bacteria often exhibit extensive phenotypic heterogeneity which arises from  stochastic intrinsic dynamics of cells. The internal state of the cell can be transmitted epigenetically in cell division, leading to correlations in the phenotypic states of cells related by descent. Therefore, a phenotypic snapshot of a collection of cells with known genealogical structure, contains information on phenotypic dynamics. Here we use a model of phenotypic dynamics on a genealogical tree to define an inference method which allows to extract an approximate probabilistic description of phenotypic dynamics based on measured correlations as a function of the degree of kinship. The approach is tested and validated on the example of Pyoverdine dynamics in {\it P. aeruginosa} colonies. Interestingly, we find that correlations among pairs and triples of distant relatives have a simple but non-trivial structure indicating that observed phenotypic dynamics on the genealogical tree is approximately conformal  - a symmetry  characteristic of critical behavior in physical systems. Proposed inference method is sufficiently general to be applied in any system where lineage information is available.
\end{abstract}

\maketitle

\section{Introduction}
Collectives of nominally isogenic cells, be it a clonal colony  of bacteria or a developing multicellular organism, are known to exhibit a great deal of phenotypic diversity and time dependent physiological variability. While often transient and reversible, phenotypic states of cells can persist on the time scale of the cell cycle and be transmitted from mother to daughter cells. This {\it epigenetic} inheritance has been a subject of much recent research and is known to involve a multitude of different molecular mechanisms \cite{Probst09,Goldberg07,Rando07}, from transcription factor transmission to DNA methylation \cite{Bird02,Riggs89}.  Stable phenotypic differentiation is at the heart of any animal and plant developmental program \cite{Li02,Hemberger09}. The role and extent of phenotypic variability in microbial populations is less well understood, but is coming into focus with the spread of single cell-resolved live imaging \cite{Locke09,Young2012} and other single-cell phenotyping methods \cite{Lubeck12}.  Phenotypic variability within a colony implements the intuitively plausible  bet-hedging strategies of survival \cite{Lachmann96,Leibler05,Raj08,Thattai,Ratcliff2010}, such as persistence \cite{Leibler04}, sporulation \cite{Veening08} or competence \cite{Caga09}. More generally, phenotypic variability may be implementing interesting ``separation of labor" -type cooperative behavior within colonies \cite{Gore13}, although evolutionary stability of such strategies remains a subject of much theoretical debate \cite{Hamilton,West07,Wilson07}. Phenotypic variation can originate from precisely controlled pattern-forming interactions either from global  or local intercellular signaling, as is the case in animal and plant development. For microbes, intracellular stochasticity is seen as playing a leading role in driving transitions between physiologically significant phenotypic states \cite{Losick08,Eldar10,Balazsi11}. It is an open problem to understand the extent to which the phenotypic diversity in a bacterial system is driven by cell-autonomous stochastic processes as opposed to the interaction with their neighbors, which could take a form of a feedback through local nutrient availability, secreted factors \cite{Waters05}, or direct contact signals \cite{Ruhe13}.

As an example, we consider {\it P. aeruginosa}, a common bacteria that like all others requires iron for metabolism, DNA synthesis, and various other enzymatic activities. To absorb iron from its naturally occurring mineral phase, {\it P. aeruginosa} produces and releases iron-chelating molecules called siderophores \cite{Hider10,Buckling07}. Pyoverdine (Pvd) is a type of siderophore that is particularly suited for experimental analysis because it is naturally fluorescent  \cite{Schalk08}. Pvd concentration varies significantly from one cell to another \cite{Nicolas}, which is largely due to the fact that Pvd is trafficking between cells that either sip or secrete them \cite{Buckling07}. Moreover, Pvd concentration along lineages has a correlation time of the order of two to three cell cycles \cite{Nicolas}. The feeder/recipient phenotypes are epigenetically passed on for a few generations before switching - a recent observation (\cite{Nicolas}) which changes the landscape of the discourse on common goods, cooperation and cheating.

Dynamics of stochastic phenotypes can be followed through multiple generations using fluorescent time-lapse microscopy and single-cell tracking \cite{Young2012,Locke09}. However, the number of distinct fluorescent reporters in a single cell is inherently limited by their spectral overlap. Alternatively, phenotypic heterogeneity can be measured with relative ease using destructive or fixed-cell methods (such as immuno-staining \cite{Gupta11} and fluorescence {\it in situ} hybridization (FISH) \cite{Lubeck12}) that only provide static snapshots. Destructive measurements can however be supplemented with lineage information (kinship) that can be collected using phase time-lapse microscopy and single-cell tracking, which is less intrusive than fluorescent microscopy. We ask: how much can one say about dynamics from a static snapshot of heterogeneity and the knowledge of the relatedness of the individuals in a population? Below we shall take a constructive approach to this question, demonstrating that by adopting a certain plausible and quite general probabilistic description of phenotypic dynamics along lineages, it is indeed possible to  infer  the dynamics from static snapshots.  We shall test the method on the  example of Pvd dynamics in {\it P. aeruginosa}, comparing the inference to direct dynamical measurements. 

Thus our goal here is to provide a tool for the study of epigenetic dynamics within proliferating collectives of cells. We shall focus on the cell-autonomous dynamics and mother-to-daughter transmission and relate the statistical description of phenotypic dynamics along any one lineage to the observable correlations between phenotypic states in a snapshot of cells at any given time, which, as we shall see, explicitly depend on the degree of kinship of the cells. Below, after framing our approach as an inference problem (Section 1 of Results), we shall define a class of models parameterizing phenotypic dynamics on lineages (Section 2 of Results) and explicitly calculate the form of ``kin correlations" from which the underlying dynamics is to be inferred. In Results Section 3, we shall apply the approach to the data on siderophore production in {\it P. aeruginosa} colonies \cite{Nicolas}, which will allow us to compare the inference results with the direct measurement of time dependent phenotypes of all cells within the colony, allowing us to validate  our approach. Last section of Results will address the question of kinship and spatial correlations within a bacterial colony. In the Discussion section we shall explain why kin correlations have a structure similar to that of correlations in conformal field theories known in physics \cite{DiFrancesco} and address possible practical applications of the approach.

\section{Results}
\subsection{Inference problem for phenotypic dynamics}
Consider a growing population of asexual individuals. At every generation, each individual gives rise to two daughters that with some probability inherit the phenotypic traits of their parent. This growing population is naturally represented as a tree (see Fig.1): the most current  population of cells corresponds to the leaves of the tree, while the branches represent its history back to the founder cell at the root. Phenotypic dynamics unfolds along the lineage linking any one leaf to the root and correlations between ``kin" arise from the fact that close relatives share more of their history. We shall assume that phenotypic dynamics is stochastic with some well defined probabilistic rule (e.g. some Markovian random process), so that the state of a cell along its lineage through the genealogical tree is a realization of the random process. 
Phenotypic variability within cell population defines a distribution of states $p_n = \frac{1}{\mathcal N}\sum_i \langle \delta_{ns_i} \rangle$ where $s_i$ is the state of cell $i$,  $\langle \hdots \rangle$ denotes averaging over the realization of the random dynamics,
${\mathcal N}$ is  the number of cells and $\delta_{ns}$ is equal to 1 if $n=s$ and 0 otherwise.  In practice, averaging over different realizations of the random process is achieved by averaging over multiple observed trees.

{\it Kin correlations} are then defined as correlations between the phenotypic states of pairs, triples or, in general, $m$-tuples of leaves with the {\it same degree} of relatedness. More specifically we characterize kin correlations by the joint distribution describing the probability of different cells to be found simultaneously in certain states. For example, for the pair-correlator, $G_{mn}^{(2)}(u)$ is defined in a given population (i.e. the single realization of the dynamics) as the fraction of all pairs of cells with the common ancestor $u$ generations in the past that are in states $m$ and $n$,
\begin{equation}
G_{mn}^{(2)}(u)= \frac{1}{\mathcal N_2} \sum_{ij, ||i-j||=u} \langle \delta_{m,s_i} \delta_{n,s_j} \rangle \label{G2mn}
\end{equation}
 $||i-j||$ is the genealogical distance, or the level of kinship, between cells $i,j$ which is defined as the number of generations to their most recent common ancestor. ${\mathcal N_2}$ is  the number of all pairs at genealogical distance $u$. Because of the possible correlations this joint probability may not be equal to the product of probabilities, $p_n p_m$, to observe $n$ and $m$ on their own. These correlations are explicitly captured by \begin{equation}
g_{mn}^{(2)}(u)= G_{mn}^{(2)}(u) -p_m p_n, \label{C2mn}
\end{equation}
that explicitly subtracts the uncorrelated (product) term.

Similarly, the triple distribution is defined by
\begin{equation}
G_{lm,n}^{(3)}(u,v)= \frac{1}{\mathcal N_3} \sum_{ijk, ||i-j||=u,||i-k||=u+v}  \langle \delta_{l,s_i} \delta_{m,s_j}\delta_{n,s_k} \rangle %>_{\{s\}} 
, \label{G3lmn}
\end{equation}
where $u$ is the number of generations to the common ancestor of the more closely related pair and $v$ is the further number of generations back to the common ancestor of all three nodes (see Fig.1), and $\mathcal N_3$ is the total number of such triplets. To focus on the correlation effects we subtract the contribution of independent fluctuations:
\begin{equation}
g_{lm,n}^{(3)}(u,v)= G_{lm,n}^{(3)}(u) -p_n g_{lm}^{(2)}(u)-p_m g_{ln}^{(2)}(v)-p_l g_{mn}^{(2)}(v)-p_l p_m p_n, \label{C3lmn}
\end{equation}
which is defined so that it goes to zero when joint probability factorizes. The so called ``connected correlators" $g_{mn}^{(2)}(u)$ and $g_{lmn}^{(3)}(u,v)$ explicitly quantify the extent of pairwise and 3rd order correlation between the nodes on the boundary of the genealogical tree. 
How much can these %connected correlators 
readily measurable correlations, defined as they are by a snapshot of a population with known genealogy, tell us about the dynamics that unfolded on the tree?

%Figure 1
\begin{figure}
\begin{center}
      \centerline{\includegraphics[scale=1]{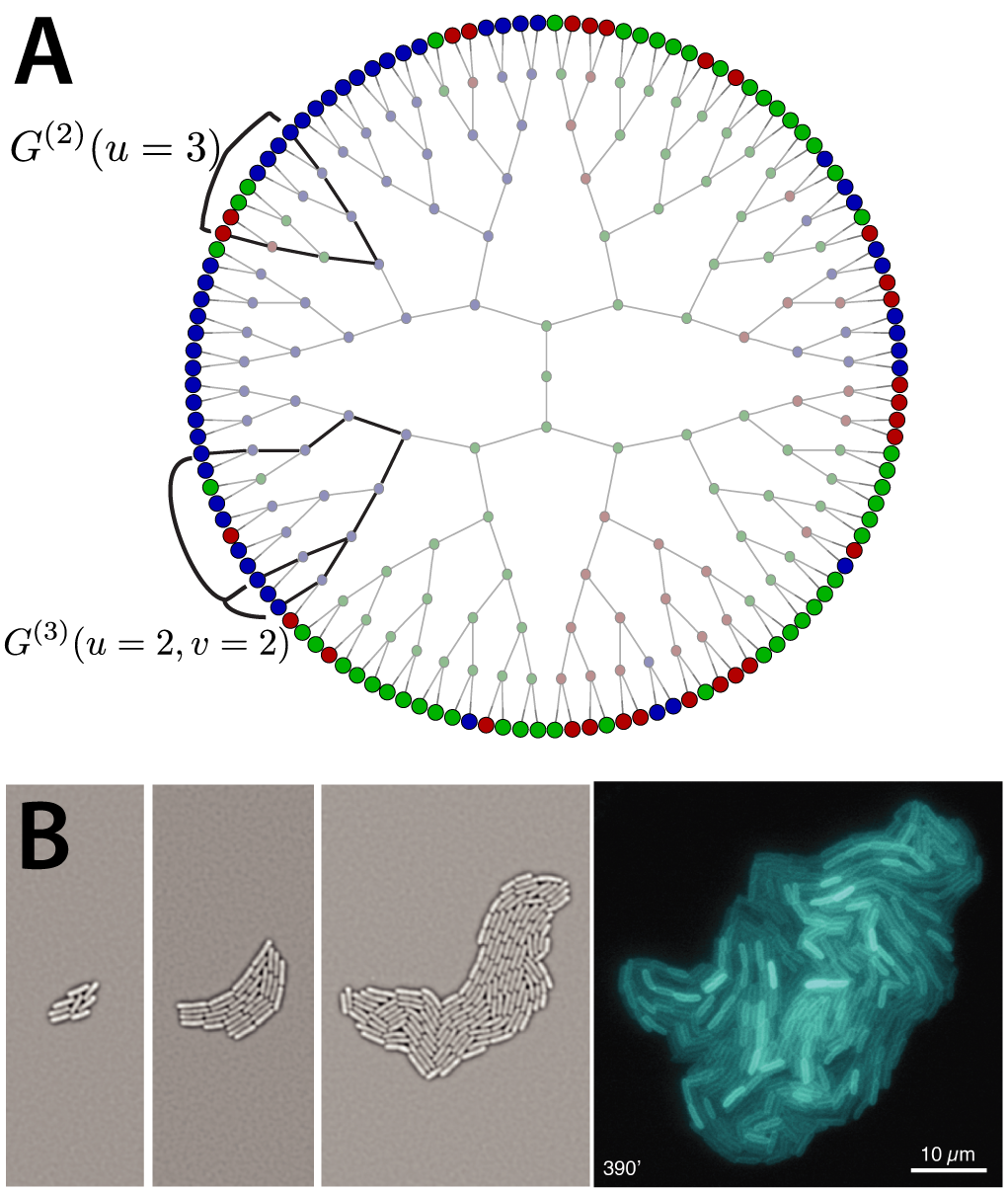}}
       \caption{(A) Phenotypic dynamics on a tree. The phenotypic state (color) of each node stochastically changes when inherited from the parent. Only the boundary nodes are accessible at the time of observation. Stochastic variation on the boundary is characterized by kin correlation functions, for example, the probability of observing blue and red colors on two nodes whose common ancestor was three generations back ($u=3$). Similarly, the three point correlation function can be investigated for a set of points, where the common ancestor of the two closest points is two generations back ($u=2$), and that of all three, an additional two generations back ($v=2$). (B) Pyoverdine distribution in a {\it P. aeruginosa} colony. Genealogy is captured by imaging the growth of the colony. At the last time-step (390') the phenotype of the bacteria --the intensity of Pvd fluorescence-- is imaged and assigned to each node.} \label{Fig1}
\end{center}
\end{figure}

%Review of Harlow et al
\subsection{The minimal model of stochastic phenotype propagation}
Let us begin with the simplest possible model. Assume that stochastic dynamics can be approximated by a Markov process, which means that probability to transition from state $n$  to a state $m$ in the time of a cell cycle depends only on the two states involved: i.e. the dynamics is defined probabilistically by a transition probability matrix $T_{1}(m|n )$ ( with $\sum_m T_1 (m|n)=1$ ). The probability $T_u (m|n)$ for a cell to start in state $n$ and end up in state $m$ time $u$ generations later is given by the product of the $u$ transition matrices obtained by iterating  $T_{u} (m|n)=\sum_k T_{1} (m|k)T_{u-1} (k|n)$. We can now calculate the joint distribution for a kin pair descending from a common ancestor in state $l$, $u$-generations back
\begin{equation}
G_{mn}^{(2)}(u)= \sum_l T_{u} (m|l)T_{u} (n|l) p_l \label{node0}
\end{equation}

The 3rd order correlator can be written down in a similar way:
\begin{equation}
\begin{split}
G_{m m', m''}^{(3)}(u,v)=  \sum_{l,k}T_{u} (m|k)T_{u} (m'|k) T_{v}(k|l)T_{u+v} (m''|l) p_l. \label{3dorde0}
\end{split}
\end{equation}

In our minimal model we assume that phenotypic states effectively form a chain with transitions occurring only between neighboring states (more generally, any graph without loops would behave the same way). 
In this case, stochastic dynamics  satisfies Detailed Balance \cite{Isihara,Harlow11} (see Supporting Information) meaning that in equilibrium the forward and backward fluxes between any pair of states balance: $T_1 (m|n)p_n = p_m T_1 (n|m)$.  This allows to define a symmetric matrix  $ A_{mn}=p_m^{-1/2}T_1 (m|n) p_n^{1/2}$ such that $A_{mn} =A_{nm}$. Note that $p_m$, being the equilibrium probability of state $m$, satisfies $\sum_n T_1(m|n) p_n=p_m$  and hence $\sum_n A_{mn} p_n^{1/2}=p_m^{1/2}$.  It is useful to diagonalize the symmetric matrix: $A \phi^{\alpha}=\lambda_{\alpha} \phi^{\alpha}$ and rewrite the transition matrix in terms of its orthonormal eigenvectors $\phi^\alpha$ and eigenvalues $\lambda_\alpha$:
\begin{equation}
T_u(m|n) = p_m^{1/2}p_n^{-1/2} \sum_{\alpha} (\lambda_{\alpha})^u \phi_m^{\alpha}\phi_n^{\alpha}   , \label{Tmn}
\end{equation}
The equilibrium distribution corresponds to the largest eigenvalue $\lambda_0=1$ and corresponding $\phi_n^{(0)} = p_n^{1/2}$.

To take full advantage of the ensuing simplifications we define correlators in the $\phi_{\alpha}$-basis:
\begin{equation}
{\hat G}_{\alpha \beta}^{(2)}(u)=  \sum_{mn} p_m^{-1/2}p_n^{-1/2} G_{mn}^{(2)}(u) \phi_m^{\alpha} \phi_n^{\beta}, \label{Ghat}
\end{equation}

In this basis, pair correlators  for our ``minimal model" of phenotypic dynamics  have a very simple form:
\begin{equation}
{\hat G}_{\alpha \beta}^{(2)}(u)=  \lambda_{\alpha}^{2u} \delta_{\alpha \beta}, \label{Ghat2}
\end{equation}
and similarly for the three-point correlator expressed in $\phi_{\alpha}$-basis in analogy with Eq.\eqref{Ghat}
\begin{equation}
{\hat G}_{\alpha \beta \gamma}^{(3)}(u,v)=  \lambda_{\alpha}^{u}\lambda_{\beta}^{u}\lambda_{\gamma}^{u+2v} C_{\alpha \beta \gamma}, \label{Ghat3}
\end{equation}
where we have defined constants:
\begin{equation}
C_{\alpha \beta \gamma} \label{C} = \sum_m p_m^{-1/2} \phi_m^{\alpha}\phi_m^{\beta}\phi_m^{\gamma}
\end{equation}
which are analogous to ``structure constants" that appear in conformal field theories describing critical phenomena in physics \cite{Itzykson89,Harlow11} - an interesting connection, which we shall explain in the Discussion section.

Note that since $\phi_m^{(0)}=p_m^{1/2}$ it follows from the orthonormality of eigenvectors $\sum_m  \phi_m^{\alpha}\phi_m^{\beta} = \delta_{\alpha \beta}$ that $C_{\alpha \beta 0} =\delta_{\alpha \beta}$. It is easy to verify that connected correlators $\hat{g}_{\alpha \beta}^{(2)}$ and $\hat{g}_{\alpha \beta \gamma}^{(3)}$ are non-zero only for $\alpha, \beta, \gamma \geq 1$ and are also given by Eqs. \eqref{Ghat2} and \eqref{Ghat3}. We emphasize that $C_{\alpha \beta \gamma}$ is determined by $\phi^\alpha$, the eigenstates of the pair correlator: thus the pair correlators fully determine the three-point correlation functions. 

In fact, it can be shown (see SI) that all of the higher order correlators can be expressed completely in terms of $\lambda_{\alpha}$ and 
$C_{\alpha \beta \gamma}$ which puts a strong and readily testable constraint on predicted correlators: pair correlator can be used to define model parameters, and higher order correlators can be used to test the model. Actually, as we shall show next, already the simple diagonal form of the expression for ${\hat G}^{(2}(u)$ is a non-trivial consequence of assumed dynamics that must be tested to verify the underlying assumptions such as existence of detailed balance.

\subsection{Kin correlations in the Pyoverdine dynamics in P. aeruginosa}
In the experiments of Julou {\it et al},  \cite{Nicolas}, the fluorescence of free Pvd in each bacterium was measured using time-lapse fluorescent microscopy, while the growth of the colony was followed with phase microscopy, providing the genealogical tree.  For the analysis below (see Methods), we used only the final snapshots of Pvd distribution for 9 colonies, each with $2^9$ cells. Each snapshot gives us Pvd  concentrations in individual cells corresponding to the leaves of a genealogical tree 9 generations deep. These concentrations  were binned to three equally likely states, denoted from $1$ to $3$, defining respectively low, medium, and high concentration states. (Connecting to the general formulation presented above, we note that in analyzing the data we can choose our freedom to define ``bins" to set $p_n$ uniform.)

%Figure Data 1
\begin{figure*}
\begin{center}
      \centerline{\includegraphics[scale=1]{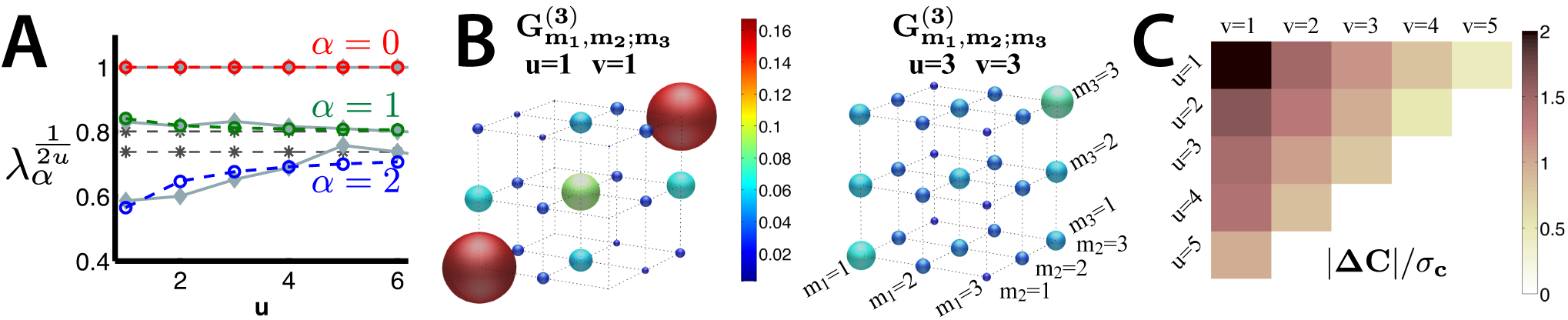}}
       \caption{Kin correlations in {\it P. aeruginosa} data. A) Eigenvalues of the two-point correlation matrix $G^{(2)}_{mn}(u)$ of the experimentally observed trees. The trivial distance dependance of the eigenvalues is scaled out (by taking the eigenvalues to the power of $1/2u$), such that the minimal model would have constant eigenvalues. Solid grey lines are the actual eigenvalues. The largest eigenvalue is always $1$ corresponding to the equilibrium mode $\alpha=0$. The dashed grey lines are the ``naive'' minimal model prediction calculated using the observed $G^{(2)}(u=5)$. The colored lines are the best fit of the interacting theory to the observed two-point correlation functions. B) The three-point correlation function $G^{(3)}_{m_1,m2_,m_3}$ computed from the data at distances $u=v=1$, and $u=v=3$. The correlation function becomes more uniform as the separation distance increases. C) Minimal model allows us to calculate the structure constants by observing only the 2nd order correlators. Deviation of the minimal model structure constants (Eq.\eqref{C}) from the actual structures constants inferred from the data, $\Delta C = C_{\alpha\beta\delta} - G^{(3)}_{obs}\lambda_\alpha^{-u-2v}\lambda_\beta^{-u}\lambda_\delta^{-u}$. The absolute value denotes the norm of matrix $\Delta C$ (methods), which is normalized by the standard deviation of the finite-size fluctuations expected in the structure constants (Methods). As expected, as the separation distance of the points on the boundary increases, the correlation functions more closely resembles those of the minimal model. The largest deviation (at $u=v=1$) has a statistical significance of two standard deviations.} \label{FigData1}
\end{center}
\end{figure*}

It is plausible to think of Pvd dynamics in the colony as a stochastic process on a tree subject to interactions that correspond to local exchange of Pvd.
We begin by comparing the observed pairwise kin correlations to the prediction of our minimal model given by Eq. \ref{Ghat2}. To that end we construct  correlation matrices for pairs of leaves conditioned by their relatedness, $u$, and diagonalize them. Fig.\ref{FigData1}A depicts the eigenvalues of the two-point correlation matrices $G_{mn}^{(2)}$ as a function of relatedness $u$. The eigenvalues are taken to the power of $1/2u$ to remove the trivial distance dependence ($\lambda_\alpha^{2u} \to \lambda_\alpha$); for the minimal model considered above, this scaling will result in eigenvalues that are independent of $u$ (see Eq.\ref{Ghat2}). The observed values, however, are significantly different from constant (see Supporting Information for the p-values), suggesting either a presence of interaction or a deviation from the simple Markovian or detailed balance form of stochastic dynamics

However, the observed eigenvalues deviate most at $u=1$ and then asymptote to a constant value with increasing $u$, suggesting the minimal model may still provide a good description of correlations among distant relatives. To test that we examined 3rd order correlators, for which the minimal model  predicts Eq.\eqref{Ghat3}: an expression defined entirely in terms of the 2nd order correlators, without any additional parameters. (As noted above, this relation is the consequence of the hidden conformal symmetry of the process.) Fig.\ref{FigData1}B,C depict the three-point correlation functions of the data. The eigenvectors of $G^{(2)}$ at $u=5$ were used as a naive approximation of $\phi_m^{\alpha}$. $\lambda_\alpha$ were approximated as eigenvalues of $G^{(2)}(u=5)$ taken to the power of $1/10$. Within statistical error, Eq.\eqref{Ghat3}, computed using $G^{(2)}$ at $u=5$, seems to correctly predict the $G^{(3)}$ at distant boundary points; however, the deviation increases as closer points on the boundary are considered. At distance $u=v=1$ the predicted 3rd order correlation function is significantly different from the experimental observation, which is not surprising, given the already noted deviations in observed pair correlations. However, the approximate agreement observed at longer genealogical distances is non-trivial and supports the validity of the model.

The fact that Eq.\eqref{Ghat3} which correctly predicts the three-point correlation function (based on the measured pair correlator) for sufficiently distant relatives demonstrates that the simple minimal model already provides a reasonable approximation for the long-time dynamics, which is quite remarkable, as it confirms approximate validity of the detailed-balance and Markovian process assumptions. We shall next demonstrate that the deviations at short times can be accounted for by existence of interaction between sisters.

%Interactions
\subsection{Effective interactions between siblings}
We now generalize our minimal model to allow for interactions between siblings, which can in effect be captured in the form of the mother-daughter transmission function $\Gamma (k_1,k_2|n)$. The two point correlator is now,

\begin{equation}
G_{mn}^{(2)}(u)= \sum_l \sum_{k_1,k_2}T_{u-1} (m|k_1)T_{u-1} (n|k_2) \Gamma (k_1,k_2 | l) p_l. \label{node}
\end{equation}
$\Gamma (k_1,k_2 | l) $ describes possible correlation in the states of the two daughters as they ``inherit" from  the mother. (Because the unconditioned effect of the mother daughter transition is subsumed in $T_{1} (m|n)$, we have $\sum_{k_2} \Gamma (k_1,k_2 | l) =T_1(k_1| l)$). Finally, $p_l$ is the probability of the ancestor to be in state $l$. 

Similarly, the 3rd order correlator is modified to:
\begin{equation}
\begin{split}
G_{m_1,m_2, m_3}^{(3)}(u,v)=  \sum_{l,k_i}T_{u-1} (m_1|k_1)T_{u-1} (m_2|k_2) \Gamma (k_1,k_2 | k_3) \\
T_{v-1}(k_3|k_4)T_{u+v-1} (m_3|k_5) \Gamma (k_4,k_5 | l) p_l. \label{3dorde}
\end{split}
\end{equation}

Without any simplifying assumptions on $\Gamma (k_1,k_2|n)$, we have a more general expression for the pair correlator:
\begin{equation}
{{\hat G}}_{\alpha \beta}^{(2)}(u)=  %\sum_{m,n} p_l^{1/2}p_m^{1/2} \phi_l^{\alpha}\phi_m^{\beta}G_{lm}^{(2)}(u)= 
\lambda_{\alpha}^{u-1}\lambda_{\beta}^{u-1}  {\hat b}_{\alpha \beta}
\end{equation}
with
${\hat b}_{\alpha \beta}=\sum_{l,m,n} {p_n \over \sqrt{p_l p_m}} {\Gamma}(l m|n) \phi_l^{\alpha} \phi_m^{\beta} $. Thanks to the consistency condition (and the fact that $\phi_m^{(0)}=p_m^{1/2}$) we have ${\hat b}_{\alpha 0}=\delta_{\alpha 0}$, so that the interaction mixes only the (decaying) $\alpha \geq 1$ eigenmodes of $T$.

In the SI we show that $\phi_m^{\alpha}$ and $\lambda_{\alpha}^u$ are still recovered as the large $u$ asymptotic eigenfunctions and eigenvalues of $G_{lm}^{(2)}(u)$. Hence they can be directly estimated from the large $u$ data. With $\phi_m^{\alpha}$ obtained by diagonalizing $G_{lm}^{(2)}(u)$ for distant kin, we can obtain ${\hat b}_{\alpha \beta}$ from the observed sister correlations ${\hat b}_{\alpha \beta}=\sum_{l,m} p_l^{-1/2}p_m^{-1/2} \phi_l^{\alpha}\phi_m^{\beta}G_{lm}^{(2)}(1) $. We can then, by diagonalizing $\lambda_{\alpha}^{u-1}\lambda_{\beta}^{u-1}  {\hat b}_{\alpha \beta}$, calculate finite $u$ corrections to $\phi_m^{\alpha}$ and $\lambda_{\alpha} $ and use these to get a corrected estimate for ${\hat b}_{\alpha \beta}$, defining an iterative process by which we fit interaction correction to the observed pair correlators.

Pair correlators however do not fully determine the $\Gamma (l,m|n)$ interaction and we next consider the 3rd order functions. Rewriting Eq. \eqref{3dorde} in terms of the eigenvectors $\phi_l^{\alpha}$ of the large $u$ (conformal) limit we find
\begin{equation}
{{\hat g}}_{\alpha \beta \gamma}^{(3)}(u,v)=  \lambda_{\alpha}^{u-1}\lambda_{\beta}^{u-1}\lambda_{\gamma}^{u+v-1}\sum_{\delta \geq 1} \lambda_{\delta}^{v-1}   \hat{\Gamma}(\alpha, \beta |\delta) {\hat b}_{\gamma \delta} \label{gInt}
\end{equation}
with the definition
\begin{equation}
\hat{\Gamma} (\alpha, \beta |\delta)=\sum_{l,m,k} \sqrt{ {p_k \over p_l p_m} } \phi_l^{\alpha}\phi_m^{\beta} \phi_k^{\delta} \Gamma(l,m|k) \label{InvGamma}
\end{equation}
which reduces to a multiple of the symmetric structure constants $\lambda_\alpha\lambda_\beta C_{\alpha\beta\delta}$ in the absence of interaction, when $\Gamma(l,m|k)= T_1(l|k)T_1(m|k)$. We also observe that  $ \hat{\Gamma} (\alpha, \beta |0)={\hat b}_{\alpha \beta} $, is already determined by the analysis of pair correlations.  Because $\lambda_{\delta}$ decreases with increasing $\delta$ one can approximate by truncating the sum over $\delta$ and  proceed to define $\hat{\Gamma} (\alpha, \beta |\delta)$ by least-square fitting the (overdetermined, on account of $u,v$ dependence) linear system relating it ${{\hat g}}_{\alpha \beta \gamma}^{(3)}(u,v)$. In practice, with limited data, we retain only the leading correction term ($\delta=1$), which as we demonstrate below can already provide a satisfactory approximation.

\subsection{Testing interaction inference on simulated data}
Above inference algorithm was applied to simulated trees with random sibling interactions $\Gamma$ (see SI for details). The empirical 2nd and 3rd order correlators were measured by counting occurrences of pairs and triplets of phenotypic states as a function of relatedness using Eqs.\eqref{G2mn} and \eqref{G3lmn}.
Eigenvectors and eigenvalues of $T_1$ ($\phi_m^{\alpha}$ and $\lambda_{\alpha}$) were calculated using the two-point correlator at the largest distance $u=6$. As discussed above, the minimal model is accurate even with interactions at large separation distances.
We then calculated a ``naive'' prediction from the minimal model of what the correlation functions ($\tilde{G}_0$) should be at other distances and higher orders using Eq. \eqref{Ghat2}-\eqref{C}. 

 %Figure 2
\begin{figure*}
\begin{center}
      \centerline{\includegraphics[scale=1]{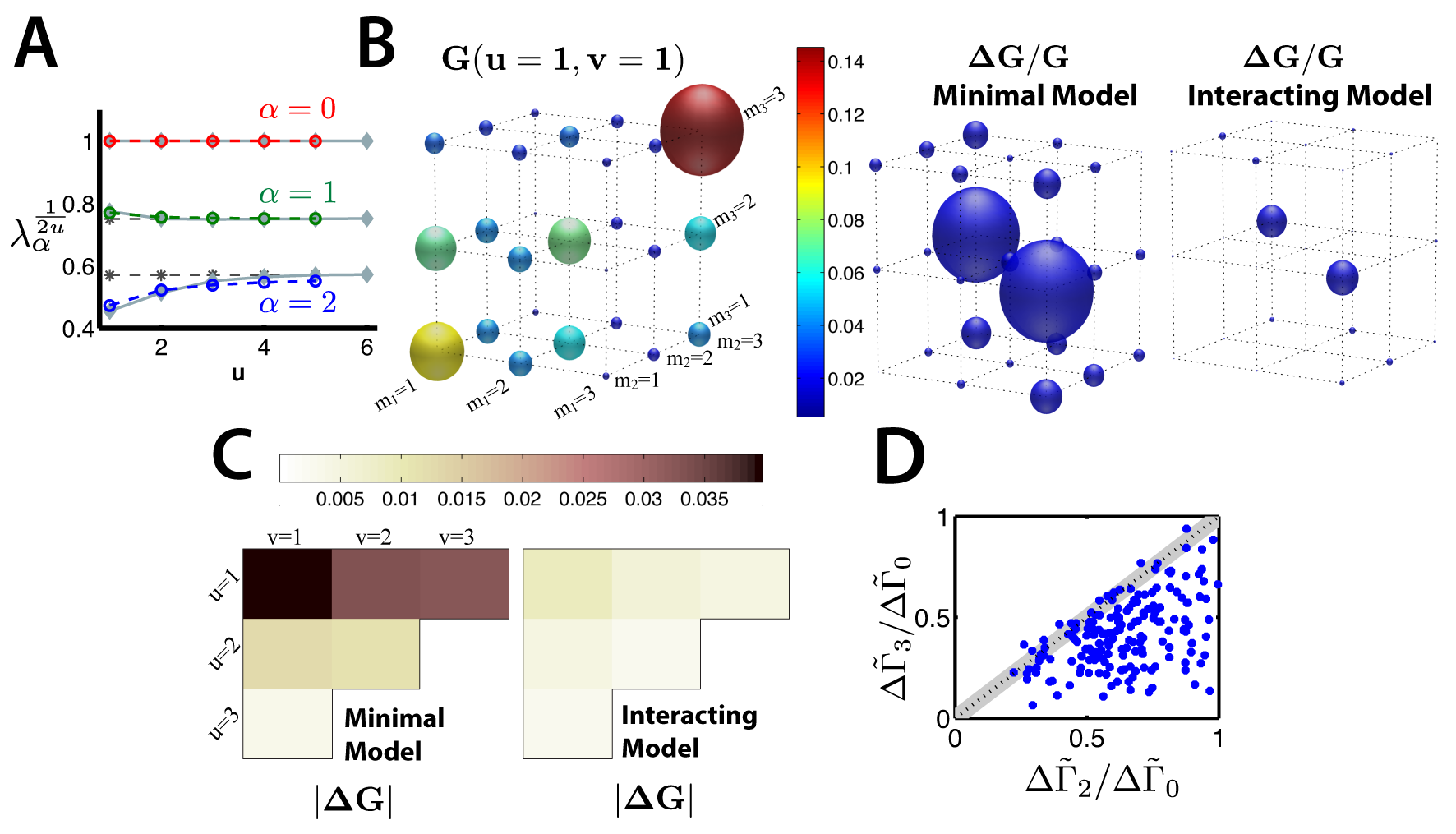}}
       \caption{{\it In silico} test of the inference algorithm. (A) Eigenvalues of the two-point correlation matrix $G^{(2)}$ of nodes whose common ancestor is $u$ generations back for trees generated from a randomly chosen interacting transition matrix $\Gamma$. The trivial scaling with distance has been removed, such that the eigenvalues of the minimal model with no sibling interactions would be constant. Solid grey lines are the actual eigenvalues computed from simulated trees with interactions (see SI for details). The largest eigenvalue is always $1$ corresponding to the stationary mode. The dashed grey lines are the minimal model prediction calculated using the observed $G^{(2)}(u=5)$. The colored lines are the interactive model fit to the observed two-point correlation functions. (B) The observed three-point correlation function $G^{(3)}(m_1,m_2,m_3)$ is depicted as $3\times3\times3$ matrix for $u=v=1$ (left). The fractional deviation of each element of the predicted three-point correlator (Eq.\eqref{Ghat3}) from its actual value, $\Delta G = G - G_{obs}$, (depicted schematically as the diameter of the spheres) for the minimal model and interacting model. The predicted correlation-functions clearly improves by introducing interactions. (C) The cumulative deviation in the predicted $G^{(3)}$ at different distances $(u,v)$ for the minimal model and with corrections from interactions. Deviation is the norm of the matrix $\Delta G$. Deviations are larger for closer boundary points. (D) Deviation of the inferred transition matrix $\tilde{\Gamma}$ from the actual one $\Gamma$ for many randomly generated $\Gamma$s. Distance based on the norm is used to quantify the deviation, $\Delta \tilde{\Gamma}_{s} = |\tilde{\Gamma}_s-\Gamma|$, for minimal model $s=0$, interactions fit to the the 2-point correlators $s=2$, and to the three-point correlators $s=3$. Fitting the three-point correlation functions improves the inference.} \label{Fig2}
\end{center}
\end{figure*}

We used three parameters to fit the deviations using the interactive form of the 2nd order correlator (corresponding to the unique non-vanishing terms in the matrix ${\hat b}_{\alpha \beta}$). Another three parameters were fit to the 3rd order correlators with the series in Eq.\eqref{gInt} terminated at the leading order $\delta = 1$ (SI). Correlators at all distances $u$ ($u,v$) were fit simultaneously to determine the free parameters in ${\hat b}_{\alpha \beta}$ (truncated $\hat{\Gamma}_1 (\alpha, \beta |\delta)$) that minimized the least-square difference between the observed and predicted ${{\hat g}}_{\alpha \beta}^{(2)}(u)$ (${{\hat g}}_{\alpha \beta \gamma}^{(3)}(u,v)$). Inferred transition matrix $\tilde{\Gamma}$ was then computed from $\hat{\Gamma}(\alpha, \beta |\delta)$ using Eq.\eqref{InvGamma}. Fig.3 shows the reduction in deviation of the predicted correlators from observed correlators as interactions are introduced into the minimal model. Fitting the three-point correlators clearly improves the inference of the transition matrix $\Gamma$ (Fig.3D).

\subsection{Inferring the interactions in P. aeruginosa}

\begin{figure*}
\begin{center}
      \centerline{\includegraphics[scale=1]{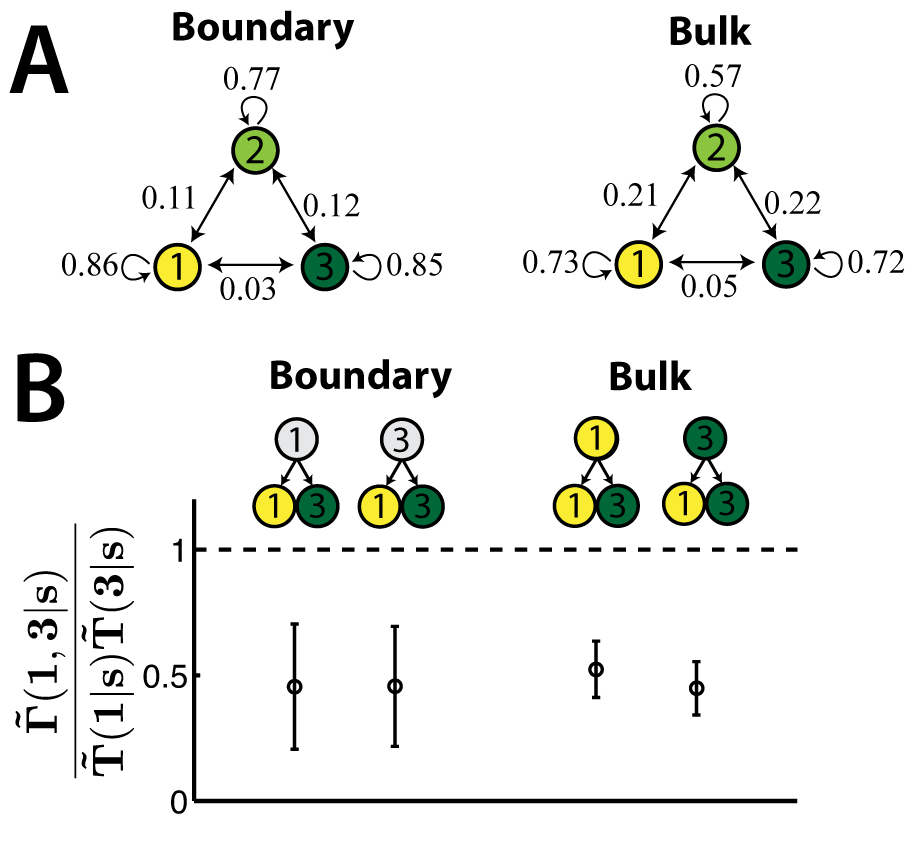}}
       \caption{Inferring the form of interactions in P. aeruginosa. A) The inferred switching rates (probability per generation) between the three Pvd states (low, medium, and high) along a single lineage, $\tilde{T}(k|s) = \sum_m \tilde{\Gamma}(m,k|s)$. $\tilde{\Gamma}$ is the inferred transition matrix by fitting the interacting model to the two-point correlation functions of the observed trees up to second-cousins. On the right, the transition rates are deduced from direct observations of Pvd states of parents and daughters in the bulk of the tree. The phenotypic state seems less likely to be conserved from direct measurements in the bulk. The discrepancy, however, is due to the ambiguity in determining the state of the parent. B) The change in likelihood of observing sibling pair 1-3 from a particular parent state $s$ due to interactions. This is the ratio of the joint-distribution of the sibling states $\tilde{\Gamma}$ over a separable distribution constructed from the marginal distributions $\tilde{T}$. The separable distribution corresponds to independent lineages with no interactions. Occurrence of 1-3 siblings is  significantly smaller with interactions. The change in likelihood is also independent of the state of the parent $s$. This is consistent with exchange interactions that decrease Pvd differentials between neighboring cells regardless of the lineage history. The inferred interactions is consistent with what is directly measured using the phenotypic states in the bulk of the tree (right). Transitions from parent state 2 to daughter states 1-3 are rare and their change in likelihood was not statistically significant in our limited data set.} \label{FigData2}
\end{center}
\end{figure*}

We now return to {\it P. aeruginosa} and attempt to infer the form of the interactions from the observed kin correlations of Pvd. Following the above recipe, we have fit the free parameters in $ {\hat b}_{\alpha \beta}$ to the observed two-point correlation functions, correctly capturing the deviations in the eigenvalues of $G^{(2)}$ matrices with $u$ (Fig.\ref{FigData1}A, colored curves). 
The inferred switching rates $\tilde{T}(k|s) = \sum_m \tilde{\Gamma}(m,k|s)$ are consistent with the switching rates that have been measured by observing the phenotypic states of parents and daughters in the bulk of the tree (Fig.\ref{FigData2}A). The apparent decrease in the probability of conserving the parental phenotype in the bulk dynamics is due to the ambiguity in determining the parent phenotype; Pvd concentration can fluctuate significantly during a cell cycle.

The inferred $\tilde{\Gamma}$ at this order --the limited nature of data does not allow us to fit higher order correlators-- contains a clear signature of interactions. The probability of one daughter cell having a low Pvd concentration while the other has a high Pvd concentration is significantly reduced compared to the non-interacting case (Fig.\ref{FigData2}B). This is consistent with nearest neighbor exchange of Pvd, which reduces sharp gradients ($1$-$3$ states) between neighboring cells, in particular siblings. Fig.\ref{FigData2}B also shows that the change in likelihood of occurrence of certain sibling pairs is independent of the state of the parent.

Moreover, from the calculated decrease in the likelihood of observing $1$-$3$ siblings pairs and the time-scale for division ($~40$ min), we can crudely estimate the exchange rate between neighbors. Define $c' = c - c_{neigh}$, the difference between Pvd concentration of a cell and its neighbor. Exchange decreases $c'$ over time, $dc'/dt = -\gamma c'$. If exchange was infinitely fast (or occurred with probability 1 at each generation) we would never observe 1-3 pairs. Our inferred interaction indicates that 1-3 pair occurs with 1/2 the frequency expected in the absence of interactions. At each generation, probability of exchange is roughly 1/2. At each exchange, $\Delta c'\sim c'$ and $\Delta t \sim 40 \times 2$ min, yielding the crude order of magnitude estimate $\gamma \sim 0.01$/min. This prediction is consistent with the value calculated from the direct observation of Pvd dynamics following individual cells \cite{Nicolas}.

\subsection{Spatial interactions}

In this section, we address the spatial nature of Pvd exchange. First, we argue that a model that only includes interactions between sisters can be used to infer interactions that take place between all neighboring bacteria in the colony. Next, we try to estimate the expected spatial correlations in Pvd concentration of cells in a colony. To do so, we restore the interactions inferred from siblings to all neighboring pairs of bacteria.

\ Local interactions in {\it P. aeruginosa} colonies are believed to be due to the exchange of Pvd. While exchange of Pvd between neighbors can correlate concentrations found in sister cells, exchange is not limited to siblings and occurs between all adjacent bacteria regardless of the degree of relatedness. Nevertheless, we shall argue that the effect of local exchange on the distribution of Pvd on the genealogical tree can be effectively represented through interactions between siblings.

\begin{figure*}
\begin{center}
      \centerline{\includegraphics[scale=1]{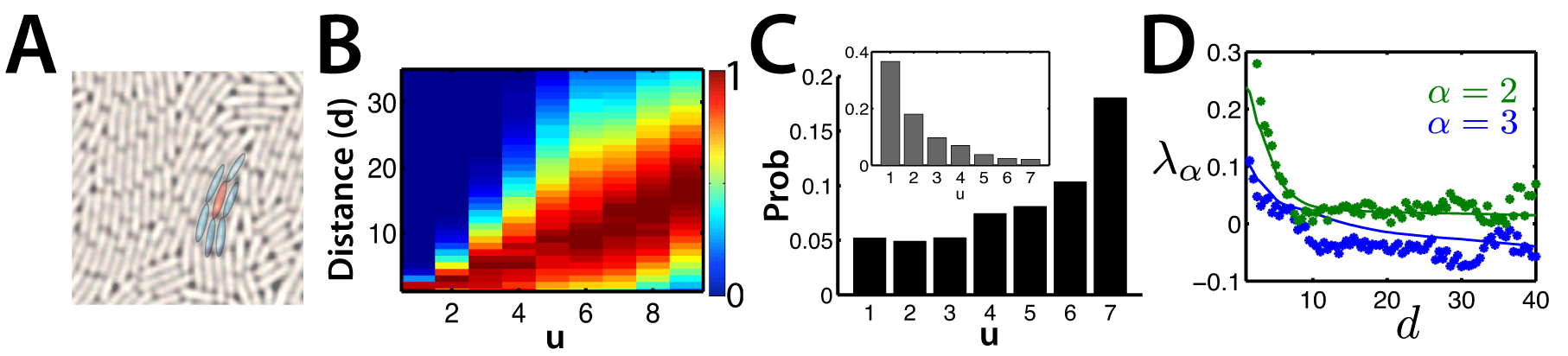}}
       \caption{Spatial proximity as a function of relatedness. A) Neighbors are defined to be within 1.5 cell widths of a given bacterium. The red bacterium above has 6 neighbors shaded in blue. The average number of neighbors is 7. B) Distribution of pair-wise spatial distances (in units of average bacterium width) between all pairs of bacteria whose common ancestor is $u$ generations back; computed over 9 colonies of 9 generations. Distribution at each value of $u$ has its maximum normalized to $1$; color scale shown on the right. The spatial distance on average increases with increasing genealogical distance. However, there are large fluctuations. C) Probability that a randomly chosen nearest neighbor pair has a common ancestor $u$ generations back. It is most likely to find distant cousins ($u=7$) adjacent to each other. This is because there are exponentially more cousins going back each generation to the common ancestor. If we normalize for number of relatives at distance $u$ (inset), we observe that with probability $\sim 0.4$ the sister will be adjacent to its sibling. However, a particular distant cousin is a neighbor with probability less than $0.03$. D) Second and third eigenvalues of the spatial correlation matrix {\bf G}$_{mn}^{(2)}(d)$ as a function of separation distance; the first eigenvalue is trivially equal to one. The solid lines are the prediction of the model (with no fitting parameters) and capture the change in correlation with decreasing separation distance.} \label{FigDistances}
\end{center}
\end{figure*}

Fig. \ref{FigDistances} shows the relationship between spatial distance and the degree of relatedness in the colonies followed in the experiments. Each bacterium has on average 7 neighbors, defined as cells located within 1.5 cells widths. Although it is more likely to find the sister as one of cell's neighbors (with probability $0.4$ compared to $0.03$ for any particular seventh-cousin), the neighborhood is dominated by distant cousins. This is because the number of cousins grows exponentially for each additional generation back to the common ancestor.

Thus, exchange with near neighbors is dominated by the exchange with distant relatives, which effectively averages over the whole distribution without contributing to kin correlations. In the limit of a well-mixed population, where neighbors are random nodes from the current generation, local exchange would contribute exactly nothing to kin correlations: any interaction that is not systematically coupled to the topology of the tree is irrelevant.  Bacteria on the plate, however are not that well mixed: the sister cell is systematically a neighbor and couples the exchange interaction to the topology of the tree. Although close relatives are also overrepresented among near neighbors, we found that to a good approximation   to account for local Pvd exchange it suffices to introduce interactions between sisters.

In the absence of direct spatial interactions, it is possible to map kin correlations to spatial correlations --the probability of observing a pair of bacteria in states $m$ and $n$ at separation distance $d$, 
\begin{equation}
\G^{(2)}_{mn}(d) = \frac{1}{\mathcal N} \sum_u G^{(2)}_{mn}(u) p(d|u)2^{u-1},
\end{equation}
where $p(d|u)$ is the probability of observing a relative of lineage distance $u$ at separation distance $d$. This distribution is determined empirically by tracking the growth of the colony and is depicted in Fig.5B. $2^{u-1}$ is the number of relatives at lineage distance $u$.

In {\it P. aeruginosa} colonies, however, direct spatial interactions exist. Local exchange of Pvd implies that kin correlations do not capture all the spatial correlations. We must reintroduce interactions between neighbors that were averaged out when we computed the kin correlators on the lineage tree. A simple way to do so is using the  following observation: progenies of distant ancestors that by chance remain nearest neighbors of closer ancestors are in effect more highly correlated than would be expected from degree of relatedness alone. This is because nearest neighbors are more likely to be in the same phenotypic state. Using this observation (see Methods) and the empirical measurement of the probability of finding a relative at lineage distance $u$ as a nearest neighbor (Fig.5C inset), we can estimate $\G^{(2)}_{mn}(d)$ without using any fitting parameters. Fig.5D shows that the prediction is in good agreement with the observed spatial correlations.

%Figure Data 2

%Figure Distances

%Discussion
\section{Discussion}

In this study we have systematically related phenotypic correlation as a function of kinship, or kin correlations, to the underlying epigenetic dynamics. Introducing a rather general class of models we were able to formulate a method for inferring dynamical parameters from static measurements on cell populations supplemented by the lineage information. This method was then applied to the data on the dynamics of Pyoverdine  in {\it P. aeruginosa} colonies with the result validated by the comparison to the direct measurements of Pvd dynamics along cell lineages.

Our analysis was based on the ``minimal model" of epigenetic dynamics which assumed 1) independent transmission  of phenotype from mother cell to its two daughters; 2) detailed balance property of  stochastic transitions between phenotypic states. The former assumption was subsequently relaxed, replaced by a  general probabilistic model of epigenetic transmission that allowed to parameterize interaction between sister cells. The profound advantage of our minimal model as a starting point is the highly constrained form of the correlations that it entails: higher order correlators are completely defined in terms of the pair-correlators. Exactly the same relation between correlators is known in field theories describing critical phenomena and is associated with conformal symmetry \cite{Harlow11,  DiFrancesco}. It is remarkable that the minimal model of epigenetic dynamics on lineages, with its highly constrained correlators, provides a good description of experimentally observed correlations among distant {\it P. aeruginosa} cells  \cite{Nicolas}.% which we found to approximately satisfy the conformal relations.

%Conformality
The relation between pair  and higher order kin correlations follows from the Detailed Balance property of the minimal model. The assumption of detailed balance in the dynamics makes forward and reverse time directions  indistinguishable: there is no ``arrow of time" associated with lineage dynamics and the tree is effectively unrooted. As a result correlations can depend only on the genealogical distance along the tree and must be explicitly independent of the position relative to the root. Now, any unrooted tree may be regarded as a finite chunk of a ``Bethe lattice", where each vertex joins three infinite binary trees, and all vertices are equivalent.

However, unlike a regular lattice (such as the square grid example in Fig.6A) where the number of nodes is a polynomial of lattice size, the Bethe lattice grows exponentially in the number of generations. A representation of the Bethe lattice where all the angles and edge lengths are constant is fundamentally impossible in Euclidean space. It is possible, however, to embed the Bethe lattice in hyperbolic space, where the negative curvature provides exponentially growing room with increasing distance \cite{Anderson05}. Fig.6B is a representation of a tree in hyperbolic space using the Poincare disk model \cite{Anderson05} (see SI for details). The angles between the edges is the same for all the nodes in this representation, and the Poincare disk metric makes all branch lengths equal.

There are transformations, such as rotation by 90 degrees and translations by integer multiples of a lattice constant, that leave the square lattice unchanged (Fig.6A). The invariance of the square lattice under these transformations implies that its correlation functions obey rotational and translational symmetries. A lattice in hyperbolic space is invariant under additional transformations. An easy way to see this is to consider the Poincare disk representation of trees (Fig.6B). Conformal transformation of the Poincare disk onto itself (see Supporting Information) are isometries that leave the lattice invariant \cite{Anderson05}. Since these transformations do not change the relative position of the bulk nodes, correlation functions on the tree must obey conformal symmetry, which accounts for their strongly constrained form \cite{DiFrancesco}.

The connection between our Eq. [9],[11] and correlators typically computed in conformal field theories \cite{Harlow11, DiFrancesco} is explained  in detail in the SI. %made more familiar by defining the distance between a pair of nodes $x$ and $y$ --with their common ancestor $u$ generations back-- as $log|x-y|=u$, and the scaling dimensions of the operators as $\Delta_{\alpha} = -log \lambda_{\alpha}$ (see SI for details). 
Yet, this unexpected connection, while providing interesting context for our findings, adds little computational power, as all of the key results followed directly from the analysis of Markovian dynamics on a tree.

%Figure 6
\begin{figure}
\begin{center}
      \centerline{\includegraphics[scale=1]{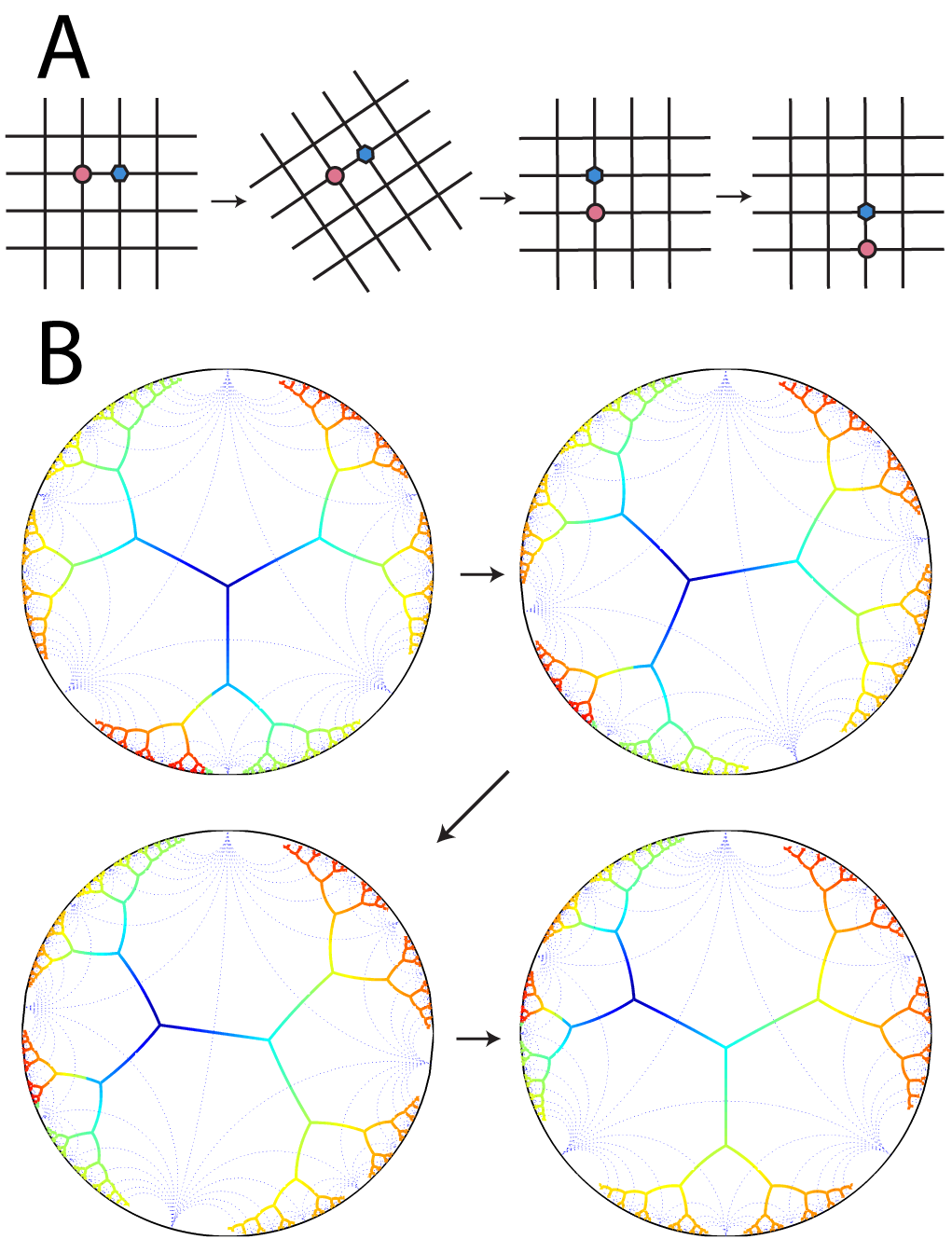}}
       \caption{Conformal symmetry of correlation functions on trees. A) There are transformations that leave the square lattice unchanged, such as rotation by 90 degrees, and translation by integer multiples of a lattice constant. Invariance of the lattice under these transformations implies that the correlation functions must obey rotational and translational symmetries. B) A tree can not be represented in Euclidean space as a lattice. However, it is possible to do so in hyperbolic space. Above, a tree in hyperbolic space is visualized using the Poincare disk; all the angles and branch lengths of the tree are constant (much like the square lattice above). The four images are snap shots of a conformal transformation that maps the tree back onto itself (the color coding is a guide for the eyes). Since the tree is unchanged, the correlation functions on a tree must obey conformal symmetry.} \label{Fig1}
\end{center}
\end{figure}

%Shortcomings
Despite its generality, the proposed approach has a number of obvious limitations. 
Virtually by definition it is blind to phenotypic dynamics that occur on the time scale shorter than a cell cycle and are not transmitted from mother to daughter. Such fluctuations do not contribute to kin correlations, furthermore, they would tend to mask the epigenetically heritable phenotypic variation. Another limitation was evident in our analysis of Pvd dynamics. Our focus on epigenetic dynamics along lineages does not allow for easy incorporation of information on spatial proximity. As a result, instead of directly estimating the interactions due to local exchange of Pvd, we estimated the {\it effect} of this interaction on kin correlation which comes about because siblings are more likely to be exchanging with each other, than with anyone else. Hence, our inference yields {\it effective} interaction, the origin of which must be examined to be properly interpreted. Other limitations of the present approach, such as discretization of the phenotypic state space state and discretization of time (corresponding to synchronously dividing population) are less fundamental. The model can be generalized to relax these assumptions if warranted by the system under consideration and the extent of available data.

%Applications
Our example of inferring Pvd dynamics should be thought of as a proof of principle. Dynamics of (naturally fluorescent) Pvd can be directly observed using time-lapse fluorescent microscopy. Dynamic reporters in general, however, require non-trivial genome engineering, and at best are limited to a few spectrally distinct fluorophores.  By contrast, measurements such as fluorescence {\it in situ} hybridization (FISH) and immuno-staining do not have these limitations, but only provide static snap shots \cite{Lubeck12,Gupta11,Kalisky11b}. High throughput technologies can simultaneously measure numerous biomarkers in large populations at a single-cell resolution \cite{Kitano02,Lubeck12,HP2,HP3,HP4,Kalisky11,Segata13,Soon13}, resulting in a snap-shot of a high-dimensional phenotypic space. Our approach is ideally suited for these applications.

%Applications more detailed
More specifically, we envision our analysis applied to understanding developmental programs and dynamics of epigenetic states in stem cells. In these systems, lineage information can be obtained from non-intrusive time-lapse microscopy, and fixed-cell measurements such as FISH can provide a snapshot of the expression levels of many genes simultaneously \cite{Singer14}. Evidence of broken detailed balance in stem cell epigenetic states can potentially shed light on the underlying pluripotency network. Similar analysis on cancer cellular states \cite{Gupta11} can elucidate the dynamics of phenotypic switching in cancer cells without a need for dynamic reporters. Moreover, lineage structure of antibody repertoires \cite{Jiang13} and tumor cells \cite{Navin11,Wang14} when supplemented with single cell phenotyping are ideally suited for analysis using our framework. Lastly, our approach can be used to disentangle phenotypic correlations due to shared lineage from those due to other factors such as signaling, which is of particular interest for understanding differentiation and reprogramming \cite{Plath11}.

%------------------------------------------------------------------------------------------------------------------------------------------------------------------------------------------------
%Methods
%% Enter any subheads and the Materials and Methods text below.
\section{Methods}

\subsection{Analyzing the experimental data}
The experimental data was in the from of a series of images captured from the growth of P. aeruginosa micro-colonies --for details of the experiments, see \cite{Nicolas}. 9 micro-colonies were analyzed. The boundary was defined to be the population on the last image. The distance of a pair of boundary nodes was calculated by counting the number of divisions from each node to their common ancestor (CA) --determined by going tracing back their history in the images. Although the division time of the bacteria was on average 40 mins, fluctuations were observed; number of generations to the CA were sometimes not the same for the two nodes. For these cases, we randomly selected the value for one of the nodes as the distance. Same method was used to determine the distances between three boundary points (values of u and v).

The signal (Pvd concentration) in each image was calculated as follows: the fluorescence intensity in the cell was subtracted from background fluorescence in that image and then normalized by the mean signal of all the cells in the image. Normalization removes the effect of increase in the total Pvd concentration in the micro-colony over time. The resultant signal distribution is stationary (see Supporting information). For the boundary cells, we discretized the signal into three phenotypes (low, medium, and high Pvd levels; respectively 1 to 3) by binning the signal to ensure a uniform distribution (equal numbers) of each phenotype.

Fig. 2C. The statistical error of the experimental data was estimated by simulating the inferred transition matrix for 64000 iterations of 9 trees of 9 generations. The eigenvalues and eigenvectors of the inferred transition matrix were obtained from the observed two-point correlation at $\smll{u=5}$. $\smll{\phi_m^\alpha}$ are the eigenvectors of $\smll{G^{(2)}(5)}$ and $\lambda_\alpha$ are its eigenvalues to the power of 1/10. $\smll{C_0}$ is calculated using the $\smll{\phi_m^\alpha}$ and Eq.\eqref{C}. $\smll{C_{obs}}$ is estimated from the 3rd order correlators using Eq.\eqref{Ghat3}. In Fig. 2C, the deviation is calculated using the matrix norm, $\smll{|C^{(3)}_{actual}(u,v) - C^{(3)}(u,v)_{0}|}$ divided by the standard deviation of $\smll{C_0(u,v)}$ over the 64000 iterations. We use Frobenius norm, which for a $\smll{N\times N}$ matrix is defined by $\smll{|A| = \sqrt{\sum_{i,j}a_{ij}^2}}$.

The bulk transition rates in Fig.4 were determined by counting all occurrences of the phenotypic states of parent and daughter cells in the observed trees. The phenotypic state of a bulk node was taken to be the Pvd state at the last time point of the cell cycle. The results were not sensitive to this choice. %The phenotypic state of the bulk nodes were discretized using the same threshold as the boundary nodes.

\subsection{Lineages in space}
Distance between any pair of bacteria in a colony is defined as the minimum distance between either pole or centroid of one bacterium to either pole or centroid of the other. Nearest neighbors are defined as pairs whose distance is less than 1.5 times the average cell width. Fig.5 was computed using spatial information from 9 colonies of 9 generations. The average coordination number is 7.

\subsection{Predicting spatial correlations}
The descendants of an ancestor at lineage distance $\smll{u'>u}$ that remain nearest neighbors of the ancestor at lineage distance $\smll{u}$ have undergone exchange with the latter ancestor for $\smll{u'-u}$ generations. These bacteria contribute to the spatial correlations not as relatives of distance $\smll{u'}$ but rather $\smll{u}$.

We include the contribution of these ``effective" ancestors as follows,
\begin{equation}
\smll{
\begin{split}
\G_{mn}^{(2)}(r) = \frac{1}{\mathcal N} \sum_u p(r|u)2^{u-1} \bigg( G_{mn}^{(2)}(u) \big(1 - \sum_{u'<u}\frac{1}{2}q(u-u') \big) +  \\ \sum_{u'>u} \big[ \theta(u'-u) G_{mn}^{(2)}(u') +   (1-\theta(u'-u))G_{mn}^{(2)}(u) \big] q(u'-u)2^{u'-u-1}  \bigg),
\end{split}
}
\end{equation}

where $\smll{\theta(\Delta u) = \frac{1}{\tau}e^{-\Delta u/\tau}}$ is the probability that exchange has not happened in $\smll{\Delta u}$ generations. $\smll{\tau}$ is the mean waiting time for exchange, which we estimated roughly as two generations (see main text), $\smll{\tau=2}$. $\smll{p(r|u)}$ is probability of finding an individual of relatedness $\smll{u}$ at spatial distance $\smll{r}$. $\smll{q(u)}$ is the empirically observed probability that a particular cousin at lineage distance $\smll{u}$ is a nearest neighbor (Fig.5C inset). $\smll{{\mathcal N}}$ is the normalization constant.

\begin{acknowledgments}
The authors thank Stephen Shenker, Douglas Stanford, Richard Neher, and David Bensimon for stimulating discussions and helpful comments. This research was supported in part by the National Science Foundation under Grant No. NSF PHY11-25915.  BIS also acknowledges support from NIH Grant R01-GM086793. ND acknowledges support from grant ANR-2011-JSV5-005-01.
\end{acknowledgments}

%%%%%%%%%%%%%%%%%%%%%%%%%%%%%%%%%%%%%%%%%%%%%%%%%%%%%%
%%%%%%%%%%%%%%%%%%%%%%%%%%%%%%%%%%%%%%%%%%%%%%%%%%%%%%

\section{Supporting Information}

%Overview
\subsection{Overview}
The Supporting Information contains the following sections. In Section 2 and 3, we discuss the operator product expansion and work out an example of a calculation of a tree correlation function. Section 4 defines the notion of connected correlators. Section 5 contains the explicit calculation of the perturbative corrections to the non-interacting minimal model when sibling interactions are added. In Section 6, we discuss the signature of broken detailed balance on the correlation functions. Section 8 deals with finite size fluctuations and presents an estimate of the number of trees required to determine statistically significant correlation functions. In the next section, we present the p-values for the deviation of the experimentally measured two-point correlation functions from the prediction of the minimal model. We also show that the distribution of Pvd states in the population is stationary. Section 9 discusses in detail the simulations used for our analysis. The last section contains our construction of the Bethe lattice in hyperbolic space.

%Example
\subsection{Example: calculating a correlation function}

It is instructive to explicitly compute a correlation function on the tree. The following example demonstrates how to introduce structure constants at the vertices and eigenvalue propagators on the edges to evaluate a correlation function (see also \cite{Harlow11}). Consider the following tree of interest --the irrelevant branches have been discarded. We would like to evaluate the correlation function of observing the four propagation modes $\alpha$, $\beta$, $\gamma$, and $\delta$ on the boundary. The input mode from the root is $0$ --the equilibrium mode with eigenvalue one.

%Figure 1
\begin{figure}
\begin{center}
      \centerline{\includegraphics[scale=1]{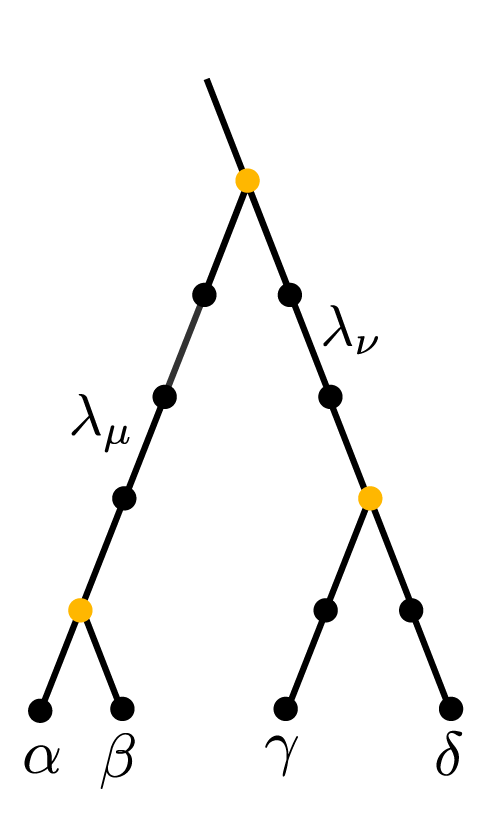}}
       \caption{Calculating correlation functions. Four points on the boundary are considered. We are interested in observing modes $\alpha$, $\beta$, $\gamma$, and $\delta$ on the boundary with the above genealogical distances. For example, $\alpha$ and $\beta$ are observed in siblings. Each edge picks up a factor $\lambda_i$, where $i$ is the mode propagating along the edge. Vertices with there edges (light color) pick up a factor of a structure constant. We sum over all possible propagation modes between internal vertices, $\mu$ and $\nu$ above. The input vector into the common ancestor of all four points is the equilibrium mode, $\alpha=0$.} \label{Fig1}
\end{center}
\end{figure}

We evaluate the correlation function as follows: For each edge insert a factor $\lambda_i$ where $i$ is the mode propagating on the edge. For each vertex with three edges insert a factor of a structure constant $C_{\alpha\beta\delta}$. Edges that connect two internal vertices correspond to the internal propagation modes ($\mu$ and $\nu$ above) which are summed over. Above example yields,
 
\begin{equation}
\hat{G}_{\alpha\beta\gamma\delta} = \lambda_\alpha \lambda_\beta \sum_\mu C_{\mu\alpha\beta} \lambda_\mu^4 \sum_\nu C_{0 \mu\nu} \lambda_\nu^3 C_{\nu\gamma\delta}  \lambda_\gamma^2 \lambda_\delta^2.
\end{equation}

Following the minimal model, we can use the diagonal form of the two point correlation functions, $C_{0 \mu \nu} = \delta_{\mu,\nu}$.

\begin{equation}
\hat{G}_{\alpha\beta\gamma\delta} = \lambda_I \lambda_H \sum_\mu C_{\mu\alpha\beta} \lambda_\mu^7  C_{\mu\gamma\delta}  \lambda_\gamma^2 \lambda_\delta^2.
\end{equation}

Above correlation functions can be converted to correlation functions in space of phenotypic states by performing a transformation.

%OPE
\subsection{Higher order correlation functions and the Operator Product Expansion}

In conformal field theories higher order correlators can be related to lower order correlators using the Operator Product Expansion (OPE) (\cite{Itzykson89}). Following Harlow {\it et al} \cite{Harlow} one can show that the same general relation holds for the Detailed Balanced Markovian dynamics on the tree. 

Consider the following n-point correlation function involving nodes $x$ and $y$ whose common ancestor is $u$ generations back.
\begin{eqnarray}
&\hat{G}^{(n)}_{\alpha\beta\gamma \hdots} =  \langle  \phi^\alpha(x) \phi^\beta(y) \phi^\gamma(z) \ldots \rangle \\
&=  \sum_\gamma C_{\alpha\beta\delta} \lambda_\alpha^u \lambda_\beta^u \lambda_\delta^{-u} \langle \phi^\delta(x) \phi^\gamma(z) \ldots \rangle \\
&= \sum_\gamma C_{\alpha\beta\delta} \lambda_\alpha^u \lambda_\beta^u \lambda_\delta^{-u}\hat{G}^{(n-1)}_{\delta \gamma\hdots}. \label{OPE1}
\end{eqnarray}

Defining the distance between nodes $x$ and $y$ as $|x-y|=e^u$ and the scaling dimensions as $\Delta_{\alpha} = -log \lambda_{\alpha}$ converts
this relation into the form

\begin{equation}
 \langle  \phi^\alpha(x) \phi^\beta(y) \phi^\gamma(z)\ldots \rangle = \sum_k \frac{C_{\alpha\beta\gamma}}{|x-y|^{\Delta_{\alpha}+\Delta_{\beta}-\Delta_{\delta}}} \langle \phi^\delta(x) \phi^\gamma(z) \ldots \rangle. \label{OPE2}
\end{equation}
which is exactly the form of OPE derived in conformal field theories \cite{Itzykson89,DiFrancesco}.

%Connected correlators
\subsection{Connected correlators}

Recall that connected pair-wise correlators are defined as
\begin{equation}
g_{mn}^{(2)}(u)= G_{mn}^{(2)}(u) -p_m p_n,
\end{equation}
where $p_m$ is the stationary probability of phenotype $m$. Similarly, the three-point connected correlators take the form,
\begin{equation}
g_{lm,n}^{(3)}(u,v)= G_{lm,n}^{(3)}(u) -p_n g_{lm}^{(2)}(u)-p_m g_{ln}^{(2)}(v)-p_l g_{mn}^{(2)}(v)-p_l p_m p_n, \label{C3lmn}
\end{equation}
Connected correlators have no contribution from independent fluctuations and decay to zero when the joint distribution of the nodes factorizes.

It is also convenient to subtract out the equilibrium component from the transition matrix, defining the ``propagator" describing approach to equilibrium
\begin{equation}
{\bf T}_u(m|n) = T_u(m|n)-p_m = p_m^{1/2}p_n^{-1/2} \sum_{\alpha=1} (\lambda_{\alpha})^u \phi_m^{\alpha}\phi_n^{\alpha}   , \label{tildeTmn}
\end{equation}
This is useful because connected correlators $g_{mn}^{(2)}$ and $g_{lmn}^{(3)}$ can be found by substituting ${\bf T}$ for $T$ in the corresponding expressions for
$G_{mn}^{(2)}$ and $G_{lmn}^{(3)}$.

%Perturbative corrections to the conformal limit
\subsection{Perturbative corrections to the conformal limit}

We calculate explicitly the leading order corrections to the conformal limit for $M=3$. In the main text, we showed that with no interactions $\tilde{\Gamma}_0(\alpha,\beta|\delta) = \lambda_\alpha\lambda_\beta C_{\alpha\beta\delta}$. We account for interactions by adding a correction to the structure constants,
\begin{equation}
\tilde{\Gamma}(\alpha,\beta|\delta) = \lambda_\alpha\lambda_\beta (C_{\alpha\beta\delta} + D_{\alpha\beta\delta}).
\end{equation}
Our goal is to fit the corrections $D_{\alpha\beta\delta}$ to the observed correlation functions.

The two-point correlation function with interactions can be represented in a matrix form.

\begin{equation}
{\mathcal C}_{\alpha\beta}(u) = \left( \begin{array}{ccc}
1 & 0 & 0 \\
0 & (1+D_{110})\lambda_1^{2u} & D_{120}\lambda_1^u\lambda_2^u \\
0 & D_{120}\lambda_1^u\lambda_2^u & (1+D_{220})\lambda_2^{2u} \end{array} \right)
\end{equation}

An observer, unaware of local interactions, would diagonalize the above matrix and infer a set of effective eigenvalues and eigenvectors. The eigenvalues are to zeroth order the minimal model eigenvalues --those of a transition matrix constructed from the marginal distribution of $\Gamma$-- plus a correction that vanishes with increasing $u$.

\begin{eqnarray*}
\tilde{\lambda}_0 &=& \lambda_0 = 1\\
\tilde{\lambda}_1 &=& \lambda_1(1+D_{110})^{\frac{1}{2u}} + \hdots = \lambda_1(1+ \frac{1}{2u} D_{110}) + \hdots \\
\tilde{\lambda}_2 &=& \lambda_2(1+D_{220} - \frac{D_{120}^2}{1+D_{110}})^{\frac{1}{2u}} + \hdots \\ &=& \lambda_2(1+ \frac{1}{2u}D_{220} - \frac{1}{2u} \frac{D_{120}^2}{1+D_{110}} ) + \hdots
\end{eqnarray*}

The eigenvectors (propagating modes) to the first order correction take the following form in the non-interacting basis
\begin{equation}
\tilde{\phi}^0 = \left( \begin{array}{c}
1  \\
0  \\
0 \end{array} \right) 
\end{equation}

\begin{equation}
\tilde{\phi}^1 = \left( \begin{array}{c}
0  \\
1  \\
\frac{D_{120}\left(\frac{\lambda_2}{\lambda_1}\right)^u}{D_{110}+1} \end{array} \right) + \hdots
\end{equation}

\begin{equation}
\tilde{\phi^3} = \left( \begin{array}{c}
0  \\
-\frac{D_{120}\left(\frac{\lambda_2}{\lambda_1}\right)^u}{D_{110}+1}   \\
1 \end{array} \right) + \hdots
\end{equation}

In the limit $u \to \infty$ (distant boundary points), the inferred eigenvalues and eigenvectors approach their non-interacting values. $u$-dependence of the eigenvectors and the scaled eigenvalues of the two-point correlation function is demonstrated using simulations in Fig.3 of the main text.

To fit the measured three-point correlators, we introduced three more parameters, $D_{111}$, $D_{121}$, and $D_{221}$. These parameters capture the leading order corrections to three-point and higher-order correlators.

%Breaking detailed balance
\subsection{Broken detailed balance}

Let us consider the general case of Markovian dynamics, which does not satisfy Detailed Balance along a single lineage:
\begin{equation}
T_u(m|n) = \sum_{\alpha} \lambda_{\alpha}^u \phi_m^{\alpha}\psi_n^{\alpha}   , \label{Tgeneral}
\end{equation}
with $\sum_n  \phi_n^{\alpha}\psi_n^{\beta} =\delta_{\alpha \beta}$. $\{\psi^\alpha \}$ form a non-orthogonal basis (assuming no degeneracies). If both $\phi$ and $\psi$ are orthonormal basis then it follows that trivially $\phi=\psi$. 
\begin{equation}
{\hat G}_{\alpha \beta}^{(2)}(u)=  \sum_{m,n}  \psi_m^{\alpha}\psi_n^{\beta}G_{mn}^{(2)}(u)= \lambda_{\alpha}^{u}\lambda_{\beta}^{u}  B_{\alpha \beta}, \label{G2general}
\end{equation}
and
\begin{equation}
{\hat G}_{\alpha \beta \gamma}^{(3)}(u,v)=  \lambda_{\alpha}^{u}\lambda_{\beta}^{u} \lambda_{\gamma}^{v} \sum_{\nu} \lambda_{\nu}^{v-u}  {\tilde C}_{\alpha \beta}^{\nu} B_{\nu \gamma}, \label{G2general}
\end{equation}
where $B_{\alpha \beta}=\sum_n p_n \psi_n^{\alpha} \psi_n^{\beta}$ and ${\tilde C}_{\alpha \beta}^{\nu}= \sum_n \phi_n^{\nu} \psi_n^{\alpha} \psi_n^{\beta}$ is the generalized ``structure constant", which is no longer fully symmetric - the fact we acknowledge by setting index $\nu$ apart from $\alpha, \beta$. (This asymmetry reflects the directionality of the dynamics in the absence of detailed balance, distinguishing the ``mother" state, here corresponding to index $\nu$,  from the daughters $\alpha, \beta$.).

An observer can infer the $\psi$ vectors and the eigenvalues $\lambda$ from observing the two-point correlation functions. However, the naive structure constants constructed from the $\psi$s in the form of the minimal model is clearly different than the actual structure constants above. The predicted three-point correlators using these structure constants would differ from the observed correlation functions.

Broken detailed balance, much like interactions, introduces `mixing' of modes. In the main text, we showed that a signature of sibling interactions was that the two point correlator matrix $\tilde{b}_{\alpha\beta}$ was no longer diagonal. Can an observer distinguish sibling interactions from broken detailed balance by observing only the two-point correlators? With enough data, the observer can infer the $\psi$ vectors and eigenvalues $\lambda$. If the $\psi$s form an orthogonal basis, the dynamics were set by sibling interactions, if not, the dynamics did not satisfy detailed balance. The difference between the two cases becomes easier to spot at higher order correlators. The structure constants for sibling interactions can be constructed form eigenvectors of two point correlators to leading order. Structure constants under broken detailed balance, however, have a $\phi$ dependence (as shown above) that can not be determined from the two point correlation functions.

%Statistical significance of the data
\subsection{Statistical analysis of the experimental data}

In the main text, we showed that the eigenvalues of the transition matrix, $\lambda_i$, computed from the two-point correlation function at various distances, $u$, deviated from the minimal model prediction. Here, we demonstrate that this deviation is statistically significant given the finite size of our data --9 trees of 9 generations.

\begin{figure*}
\begin{center}
      \centerline{\includegraphics[scale=1]{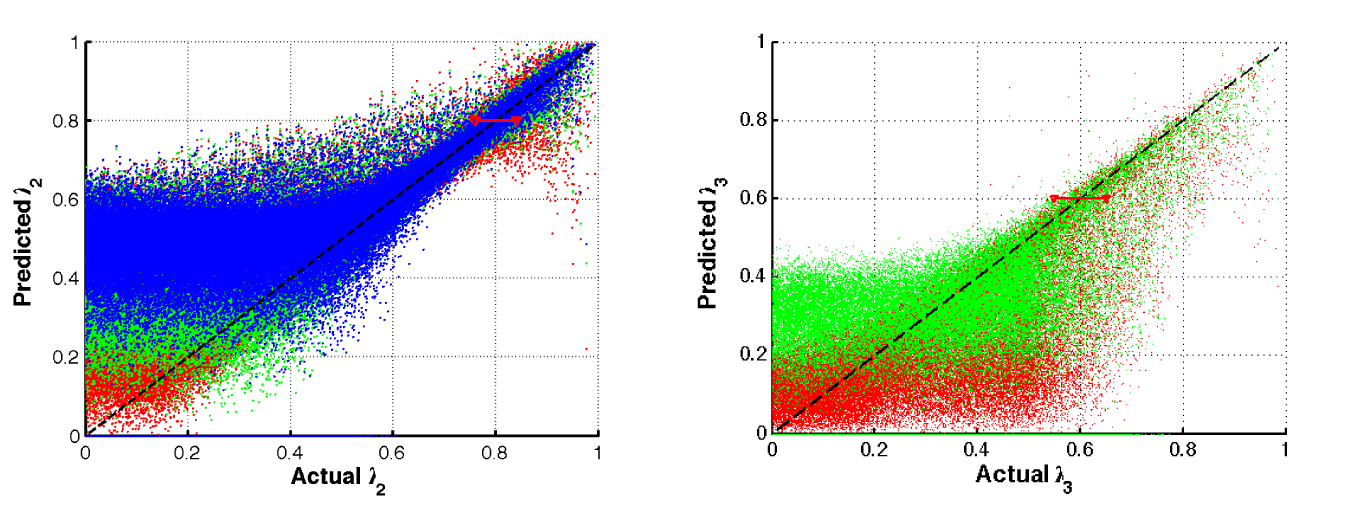}}
       \caption{Simulated trees with random transition matrices. Predicted versus actual eigenvalues (second and third; left and right respectively) for simulated trees with random transition matrices that satisfy detailed balance and have no interactions. The colors denote the distance $u$ at which the two-point correlation function is used to estimate the eigenvalues; blue, green, and red, correspond to $u=1$ to $u=3$ respectively. The red arrows show the eigenvalues estimated from the data. The small eigenvalues have almost no predictive power since the amplitude of their corresponding propagation mode is smaller than the statistical fluctuations.} \label{FigSimEigs}
\end{center}
\end{figure*}

First, we checked whether eigenvalues computed from the two-point correlation functions had predictive power given the statistical fluctuations in a finite data set. We simulated many trees of the same size as the data for random transition matrices of size 3. Fig.\ref{FigSimEigs} depicts the predicted eigenvalues versus the actual ones for two-point correlation functions at different distances. The first eigenvalue is trivially one; only the second and third eigenvalues are shown. The predictions correlate well with the actual values when the eigenvalues are large. A large eigenvalue corresponds to a long-lived fluctuation mode --see main text-- and is easier to detect. Fortunately, the eigenvalues estimated from the correlation functions in the data are large enough (roughly 1, 0.8, and 0.6 --see figure) to fall in the region that the predictions are well-correlated with the actual values.

%Figure SimEigs

Fig.2A in the main text depicts the departure in the measured two-point correlation functions from the prediction of the minimal model. However, deviations are expected simply due to the statistical fluctuations in the correlation functions from finite size effects. To quantify the statistical significance of the observed deviation, we simulated 10000 iterations of 9 trees of 9 generations using a transition matrix deduced from the observed two-point correlation function at $u=6$ (the starting point of the minimal model) and with no interactions. Fig.\ref{FigSimEigs} shows the distribution of the third eigenvalue of the resulting two-point correlation function of these trees for different distances. The observed values from data are also shown as red-arrows. The p-values show that the deviation is statistically significant for distances $u=1$ to $u=3$.

%Figure DatapVal
\begin{figure*}
\begin{center}
      \centerline{\includegraphics[scale=1.2]{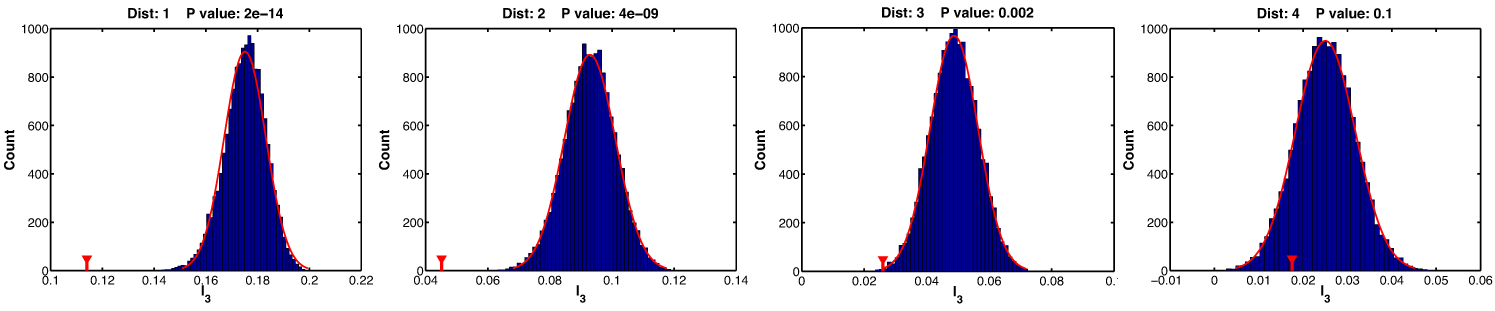}}
       \caption{Statistical significance of the deviation in the observed two-point correlation functions from that of the null-hypothesis. Top) Histogram of the third eigenvalue of the two-point correlation matrix of simulated trees with the same transition matrix as the null-hypothesis minimal model, for distances $u=1$ to $u=4$, from left to right. The red arrow shows the value from the measured two-point correlation function of the data. The p-values are shown on the top.} \label{FigDatapVal}
\end{center}
\end{figure*}

\subsection{Finite size fluctuations}

As discussed above, the finite size of the data imposes statistical limits on the accuracy of the correlation functions. Here, we roughly estimate the number of cells that must be observed for the correlation functions to be statistically significant. Assume that $n_T$ trees of $u$ generations are observed each containing ${\mathcal N} = 2^u$ cells. We would like to use this observation to compute the $q$-th order correlation function $G^{(q)}$.

$G^{(q)}$ is constructed from q-tuples of the leaves on the tree, total number of which is given by ${{\mathcal N} \choose q}<\big(\frac{e{\mathcal N}}{q}\big)^q$. $q$-th order correlator is a matrix with at most $M^q$ independent entires, where $M$ is the total number of states (e.g. $M=3$ in our analysis of Pvd). The correlation is also a function of at least $q-1$ distances, each of which can vary from $1$ to $u$. A conservative estimate of the number of observations for a single element of $G^{(q)}$ at a given distance is $n_Te^q{\mathcal N}^q/(M^qq^qu^{q-1})$.

The fluctuations in the computed correlator is roughly the shot noise in the number of observations. We set this equal to the theoretical estimate of the correlator set by the transition matrix $T$. Denote the smallest eigenvalue of $T$ as $\lambda$. The smallest theoretical correlation (frequency of observing a certain q-tuple) is roughly $\lambda^{qu}$. Setting this equal to the shot noise in the finite number of observations, $n_T^{1/q} \sim  qM/e(\sqrt{2}\lambda)^{2u} $. The number of required trees grows exponentially with the degree of the correlation function.

\subsection{Analysis of Pvd signal in the bulk of the tree}

We had access to the Pvd signal in the bulk of the tree from time-lapse fluorescent microscopy during the growth of the colonies. As discussed in the main text, the bulk dynamics of Pvd is more complex than the effective theory constructed using the boundary nodes. For instance, the Pvd concentration in a given bacterium can change significantly between divisions. Nevertheless, we set out to confirm the key attributes of the bulk dynamics that must hold for our approach to be valid. We considered fluorescence of Pvd from images taken at earlier time points (frames). Same normalization was used as for the boundary nodes (see Methods in the main text), but involving only the cells in the given frame. Fig.\ref{FigBulkSignal} shows the distribution of Pvd in the micro-colony for different time points. The distribution is stationary consistent with existence of detailed balance. 

%Figure BulkSignal
\begin{figure}
\begin{center}
      \centerline{\includegraphics[scale=0.6]{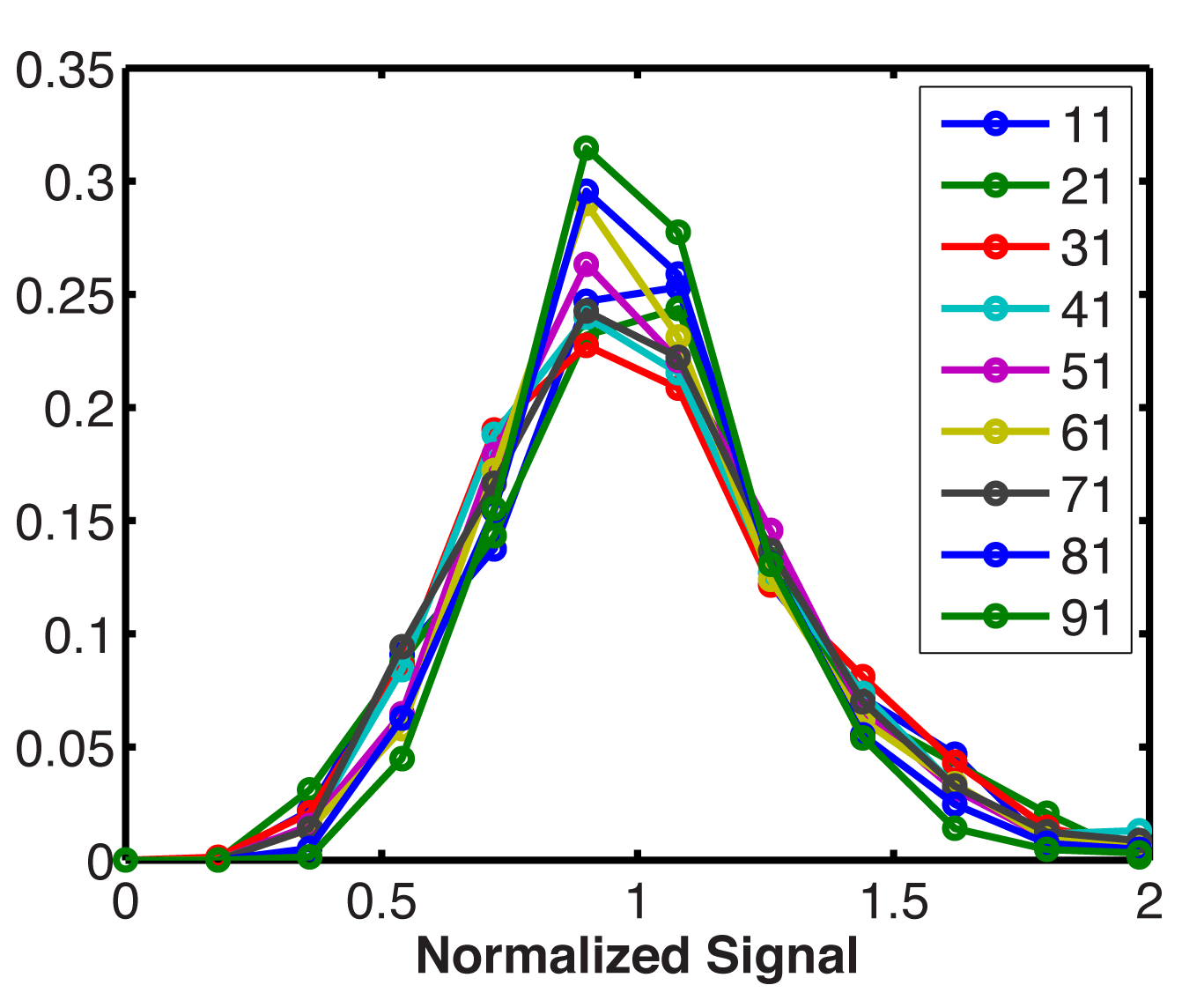}}
       \caption{Distribution of the Pvd signal of all cells in a given frame (time-point) averaged over 10 adjacent frames and all 9 trees of the data. The legend shows the frame-number. Each generation is roughly 10 frames. The distribution is stationary.} \label{FigBulkSignal}
\end{center}
\end{figure}

%Simulations
\subsection{Simulations} 
We simulated 320000 trees of size 6 generations using three states (phenotypes), M=3. The transition matrix ${T(m|n)}$ is a randomly chosen symmetric ${3\times3}$ matrix whose columns add to 1. We also required that the minimum eigenvalue of T be larger than 0.5 to ensure long-lasting fluctuation modes. For each tree the root was uniformly drawn number from 1 to N. For non-interacting trees, at each generation, every node gives rise to two off-springs. Each off-spring is independently assigned color ${m}$ with probability ${P(m|n)}$, where n is the color of the parent.

For the simulations, a random transition matrix with sibling interactions  ${\Gamma(m_1,m_2|n_d)}$ --a ${3\times3\times3}$ matrix-- was generated. We did so by starting from a non-interacting version constructed from the  randomly generated marginal distribution ${T(m|n)}$ (see above), ${\Gamma_0(m_1,m_2|n_d) = T(m_1|n_d)T(m_2|n_d)}$. An inseparable distribution was created by adding a noise term to each element of the matrix (drawn from a Gaussian distribution with mean 0 and standard deviation 0.01), in such a way that the marginal distribution was not changed, and the symmetry condition satisfied, ${\Gamma(m_1,m_2|n_d)=\Gamma(m_2,m_1|n_d)}$.

Two-point correlation functions ${G^{(2)}_{n_a,n_b}(u)}$ was computed by counting all pairs of the final generation with colors ${n_a}$ and ${n_b}$ whose common ancestor is ${u}$ generations back and dividing by total number of such pairs. The ordering of the pair is discarded ensuring that matrix ${G^{(2)}_{n_a,n_b}(u)}$ is symmetric. Three-point correlation function ${G^{(3)}_{n_a,n_b,n_c}(u,v)}$ was similarly computed. The order of the closest pair is discarded ensuring that  ${G^{(3)}}$ is symmetric in indices ${n_b}$ and ${n_c}$.

In Fig.3C (main text), we plot the difference between the predicted minimal model three-point correlators (computed using Eqs.10,11 (main text) with ${\phi_m^\alpha}$ and ${\lambda_\alpha}$ equal to eigenvectors of ${G^{(2)}(u=6)}$ and its eigenvalues to the power of 1/12 respectively) and the the measured three-point correlators. For each point ${u}$ and ${v}$, we have plotted the norm of the difference of the two matrices. For the interacting model, the predicted correlation functions are determined by first fitting the 3 ${\tilde{b}_{\alpha}{\beta}}$ to the observed two-point correlators, and three ${\tilde{\Gamma}(\alpha,\beta|1)}$ parameters to the observed three-point correlators.

The parameters of the interacting theory were fit to the simulated/experimental data using an unconstrained nonlinear optimizer implemented using the line search algorithm \cite{Fletcher87}, programmed in Matlab R2011.  In Fig. 4B (main text), the deviation between the predicted interaction matrix and the actual one is defined as: ${\Delta \tilde{\Gamma}^{(s)} = |\tilde{\Gamma}^{(s)} -  \Gamma|_F}$, where s=0,1,2 denotes respectively the prediction of the minimal model, fit to the 2nd order correlators, and the fit to 2nd and 3rd order correlators.

%Hyperbolic trees
\subsection{Hyperbolic trees}

\begin{figure*}
\begin{center}
      \centerline{\includegraphics[scale=0.8]{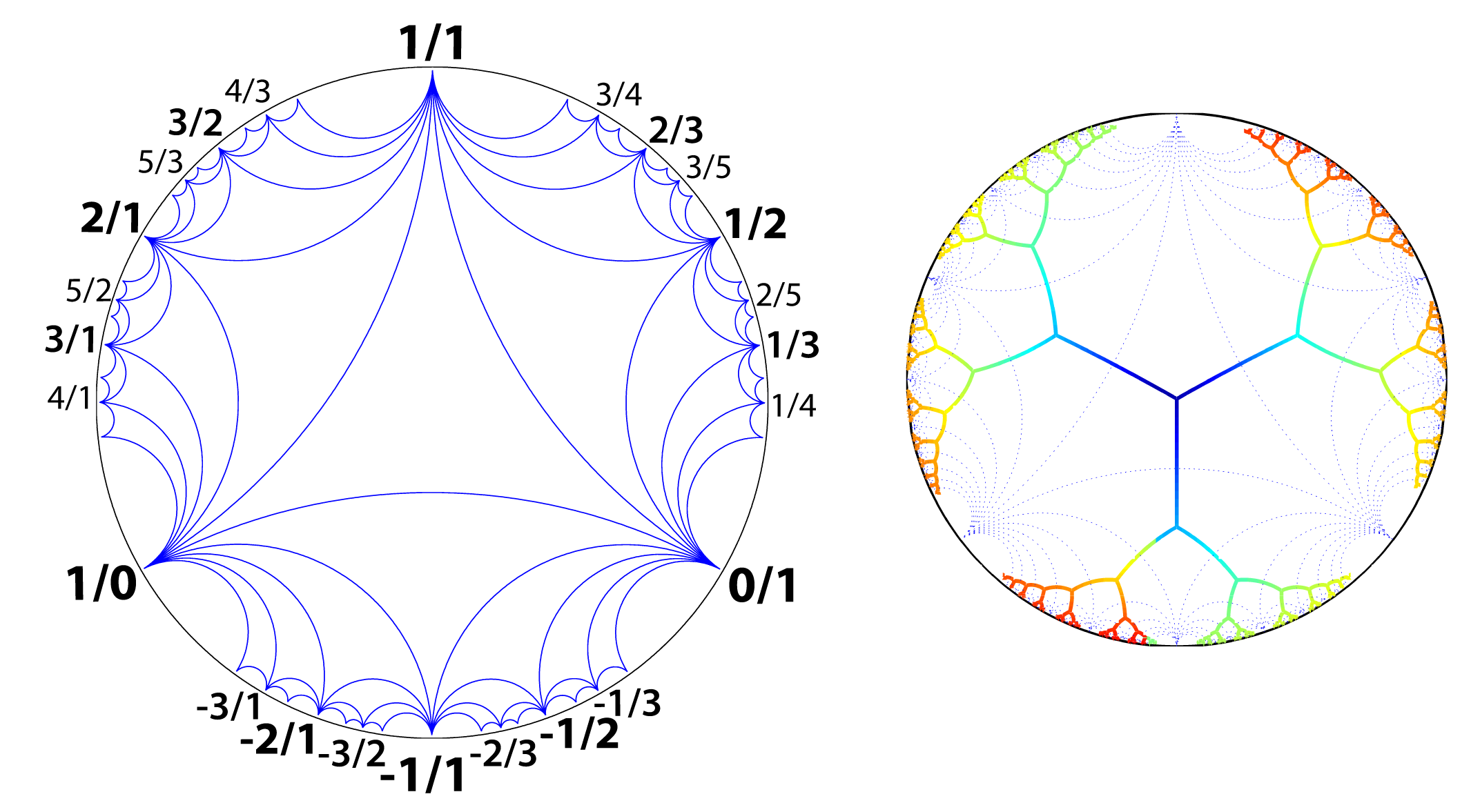}}
       \caption{Bethe lattice in the Poincare disk. (Left) Triangulation of the hyperbolic space in the Poincare disk representation using the Farey sequence construction. The fractions show how the Farey sequence divides up the unit circle boundary into smaller segments (this is a mapping of the real line to the boundary of the disk). Size of the font is reduced at each consecutive iteration of the Farey construct. (Right) The Bethe lattice is the dual to the triangulation and constructed by drawing the geodesics connecting the centroids of neighboring triangles. The coloring scheme is a guide for the eyes.} \label{FigBulkSignal}
\end{center}
\end{figure*}

We use the standard half-plane and Poincare disk models as representations of hyperbolic space (see for example \cite{Anderson05}). The Bethe lattice in hyperbolic space (Fig.6 in the main text) is constructed by creating a triangulation of the half-plane model first. The vertices of the triangulation are constructed using the Farey sequence as follows: start with the set of fractions $\{ 0/1,1/0\}$; at each iteration, add the mediant of each consecutive pair of fractions to the set (mediant of fractions $\frac{a}{b}$ and $\frac{c}{d}$ is $\frac{a+b}{c+d}$). The first few iterations starting with $\{ 0/1,1/0 \}$ result in sets:  $\{ 0/1,1/1,1/0\}$, $\{ 0/1,1/2,1/1,2/1,1/0\}$, $\{ 0/1,1/3,1/2,2/3,1/1,3/2,2/1,3/1,1/0\}$, etc. Each new element in the set is a new vertex (on the real line) that is connected by geodesics in the half-plane to its two parent vertices. The Bethe lattice is the dual of the triangulation constructed by drawing the geodesics connecting the centroids of neighboring triangles.

The resulting triangulation and Bethe lattice in the half-plane is mapped onto the Poincare disk using the standard transformation \cite{Anderson05} (see Figure 5 in SI),
\begin{equation}
 z \to \frac{z-i}{iz-1}.
\end{equation}
Transformations of the following form (known as Mobius transformations) are isometries of hyperbolic space and map the Poincare disk onto itself \cite{Anderson05}
\begin{equation}
 z \to \frac{\alpha z + \beta}{\bar{\beta}z + \bar{\alpha}},
\end{equation}
where $\alpha,\beta \in \mathbb{C}$ and $|\alpha|^2-|\beta|^2 = 1$. For the transformation shown in Fig.6B of the main text, we repeatedly applied the infinitesimal transformation, $\alpha = 1 - i\epsilon$ and $\beta = -\epsilon$. This transformation maps the tree onto itself but effectively shifts the origin by one node. We used the same transformation on the Farey triangulation to move one node to the origin of the Poincare disk, obtaining the 3-fold symmetry. Designating a particular node as the origin is an arbitrary choice with no bearing on the correlation functions. Therefore, correlation functions on a tree are invariant under above transformation.

%%%%% References
%\bibliographystyle{abbrv}
%\bibliography{main}

\begin{thebibliography}{10}

%Epigenetics
\bibitem{Probst09} Probst AV, Dunleavy E, Almouzni G (2009) Epigenetic inheritance during the cell cycle. Nat Rev Mol Cell Biol 10:192-206.

\bibitem{Goldberg07} Goldberg AD, Allis CD, Bernstein E (2007) Epigenetics: A landscape takes shape. Cell 128(4):635-638.

\bibitem{Rando07} Rando OJ, Verstrepen KJ (2007) Timescales of genetic and epigenetic inheritance. Cell 128:655Ð668.

%Methylation
\bibitem{Bird02} Bird A (2002) DNA methylation patterns and epigenetic memory. Genes Dev 16:6-21.
\bibitem{Riggs89} Riggs AD (1989) DNA methylation and cell memory. Cell Biophys 15:1-13
%Development Histone modification
\bibitem{Li02} Li E (2002) Chromatin modification and epigenetic reprogramming in mammalian development. Nat Rev Genet 3(9):662-673. 
%Development
\bibitem{Hemberger09} Hemberger M, Dean W, Reik W (2009) Epigenetic dynamics of stem cells and cell lineage commitment: Digging Waddington's canal. Nat Rev Mol Cell Biol 10:526-537.



%Single cell characterization methods
%Elowitz Time Lapse Microscopy
\bibitem{Locke09} Locke JCW, Elowitz MB (2009) Using movies to analyse gene circuit dynamics in single cells. Nat Rev Microbiol 7:383-392.

\bibitem{Young2012} Young JW, et al. (2012) Measuring single-cell gene expression dynamics in bacteria using fluorescence time-lapse microscopy. Nat Protoc 7(1):80-88.

%FISH
\bibitem{Lubeck12} Lubeck E, Cai L (2012) Single-cell systems biology by super-resolution imaging and combinatorial labeling. Nat Methods 9(7):743-748.


\bibitem{Lachmann96} Lachmann M, Jablonka E (1996) The inheritance of phenotypes: An adaptation to fluctuating environments. J Theor Biol 181:1-9.

\bibitem{Leibler05} Kussell E, Leibler S (2005) Phenotypic diversity, population growth, and information in fluctuating environments. Science 309:2075-2078.

\bibitem{Raj08} Raj A, van Oudenaarden A (2008) Nature, nurture, or chance: Stochastic gene expression and its consequences. Cell 135:216-226.

%Bet-hedging
\bibitem{Thattai} Thattai, M \& van Oudenaarden, A (2004) Stochastic gene expression in  fluctuating environments. Genetics 167, 523-530.
\bibitem{Ratcliff2010} Ratcliff, WC \& Denison, RF (2010) Individual-level bet hedging in the bacterium Sinorhizobium meliloti. Curr. Biol. 20, 1740-1744.

%Phenotypic heterogeneity is functional
\bibitem{Leibler04} Balaban NQ, Merrin J, Chait R, Kowalik L, Leibler S (2004) Bacterial persistence as a phenotypic switch. Science 305, 1622.


%Sporulation
\bibitem{Veening08} Veening JW, et al. (2008) Bet-hedging and epigenetic inheritance in bacterial cell development. Proc Natl Acad Sci USA 105(11):4393-4398.


%Competence
\bibitem{Caga09} Caatay T, Turcotte M, Elowitz MB, Garcia-Ojalvo J, SŸel GM (2009) Architecture-dependent noise discriminates functionally analogous differentiation circuits. Cell 139:512-522. 

%Microbial cooperation
\bibitem{Gore13} Celiker H, Gore J (2013) Cellular cooperation: Insights from microbes. Trends Cbell Biol 23(1):9-15. 

%Theoretical debate evolution of cooperation
\bibitem{Hamilton} Hamilton WD (1964) The genetical evolution of social behaviour. I. J Theor Biol 7:1-16.
\bibitem{West07} West SA, Griffin AS, Gardner A (2007) Evolutionary explanations for cooperation. Curr Biol 17:R661-R672.
\bibitem{Wilson07} Wilson DS, Wilson EO (2007) Rethinking the theoretical foundation of sociobiology. Q Rev Biol 82(4):327-348.

\bibitem{Losick08} Losick R, Desplan C (2008) Stochasticity and cell fate. Science 320:65-68.

\bibitem{Eldar10} Eldar A, Elowitz MB (2010) Functional roles for noise in genetic circuits. Nature 467:167, 173.

\bibitem{Balazsi11} Bal‡zsi G, van Oudenaarden A, Collins James J (2011) Cellular decision making and biological noise: From microbes to mammals. Cell 144:910Ð925.

%Bacterial signalling
\bibitem{Waters05} Waters CM, Bassler BL (2005) Quorum sensing: Cell-to-cell communication in bacteria. Annu Rev Cell Dev Biol 21:319-346.
\bibitem{Ruhe13} Ruhe ZC, Low DA, Hayes CS (2013) Bacterial contact-dependent growth inhibition. Trends Microbiol 21(5): 230-7.


%Siderophore
\bibitem{Hider10} Hider RC, Kong X (2010) Chemistry and biology of siderophores. Nat Prod Rep 27(5): 637-657.

%Siderophore
\bibitem{Buckling07} Buckling A, et al. (2007) Siderophore-mediated cooperation and virulence in pseudomonas aeruginosa. FEMS Microbiology Ecology 62:135-141.

%Siderophore
\bibitem{Schalk08} Schalk IJ (2008) Metal trafÞcking via siderophores in gram-negative bacteria: SpeciÞcities and characteristics of the pyoverdine pathway.J Inorg Biochem 102(5-6):1159-1169.


\bibitem{Nicolas} Julou T, Mora T, Guillon L, Croquette V, Schalk IJ, Bensimon D, Desprat N (2013) CellÐcell contacts confine public goods diffusion inside Pseudomonas aeruginosa clonal microcolonies. PNAS 110:12577-12582.


%Immuno staining
\bibitem{Gupta11} Gupta PB, et al. (2011) Stochastic state transitions give rise to phenotypic equilibrium in populations of cancer cells. Cell 146:633-644. 

\bibitem{DiFrancesco} DiFrancesco P, Mathieu P, Senechal D (1999) Conformal Field Theory (Springer, New York).

\bibitem{Isihara} Isihara A, Statistical Physics (Academic Press, NY 1971).

\bibitem{Harlow11} Harlow D, Shenker S, Stanford D, Susskind L (2011) Eternal Symmetree. arXiv:1110.0496 [hep-th].

\bibitem{Itzykson89} Itzykson C, Drouffe JM (1989) Statistical Field Theory (Cambridge Univ Press, Cambridge, UK).



%\bibitem{Koblitz} Koblitz N (1977) p-Adic Numbers, p-adic Analysis and Zeta Functions, Graduate Texts in Mathematics Vol. 58 (Springer-Verlag, Berlin).

%Hyperbolic geometry
\bibitem{Anderson05} Anderson JW (2005) Hyperbolic Geometry (Springer-Verlag, London).


%\bibitem{Moxon94} Moxon ER, Rainey PB, Nowak MA, Lenski RE (1994) Adaptive evolution of highly mutable loci in pathogenic bacteria. Curr. Biol. 4:24.

%\bibitem{Kearns10} Kearns DB (2010) A field guide to bacterial swarming motility. Nat Rev Microbiol 8:634-644.

%\bibitem{Smith02} Smith CM, et al. (2002) Heritable chromatin structure: mapping ÒmemoryÓ in histones H3 and H4. Proc Natl Acad Sci USA 99:16454-16461.

%\bibitem{Li00} Li L, Lindquist S (2000) Creating a protein-based element of inheritance. Science 287:661-664.

%\bibitem{Corson12} Corson F, Siggia ED (2012) Geometry, epistasis, and developmental patterning. Proc Natl Acad Sci USA, 109:5568-5575.


%High through-put
\bibitem{Kitano02} Kitano H (2002) Systems biology: A brief overview. Science 295:1662Ð1664.

\bibitem{HP2} Loo LH, Wu LF, Altschuler SJ (2007) Image-based multivariate profiling of drug responses from single cells. Nat Methods 4:445-453.

\bibitem{HP3} Rauch T, Pfeifer GP (2005) Methylated-CpG island recovery assay: A new technique for the rapid detection of methylated-CpG islands in cancer. Lab Invest 85:1172Ð1180.

\bibitem{HP4}  Bibikova M, et al. (2006) High-throughput DNA methylation profiling using universal bead arrays. Genome Res 16:383Ð393. 


\bibitem{Kalisky11} Kalisky T, Blainey P, Quake SR (2011). Genomic analysis at the single-cell level. Annu. Rev. Genet. 45, 431-445.

\bibitem{Segata13} Segata N, et al. (2013) Computational metaÕomics for microbial community studies. Molecular Systems Biology 9:666.

\bibitem{Soon13} Soon WW, Hariharan M, Snyder MP (2013) High-throughput sequencing for biology and medicine. Molecular Systems Biology 9: 640.

\bibitem{Singer14} Singer ZS, et al. (2014) Dynamic heterogeneity and DNA methylation in embryonic stem cells. Mol. Cell 55: 319-331.

\bibitem{Jiang13} Jiang N, et al. (2013) Lineage structure of the human antibody repertoire in response to influenza vaccination. Sci Transl Med 5(171):171ra119.

\bibitem{Navin11} Navin N, et al. (2011) Tumour evolution inferred by single-cell sequencing. Nature 472(7341):90-94.

\bibitem{Wang14} Wang Y, et al. (2014) Clonal evolution in breast cancer revealed by single nucleus genome sequencing. Nature 512(7513):155-160.

\bibitem{Plath11} Plath K, Lowry WE (2011) Progress in understanding reprogramming to the induced pluripotent state. Nat Rev Genet 12(4):253-265.

%\bibitem{Halabi09} Halabi N, Rivoire O, Leibler S, Ranganathan R (2009) Protein sectors: Evolutionary units of three-dimensional structure. Cell 138:774-786.

%Single cell sequencing
\bibitem{Kalisky11b} Kalisky T, Quake SR (2011) Single-cell genomics. Nat Methods 8(4):311-314.

\bibitem{Fletcher87} Fletcher R (1987) Practical Methods of Optimization  (John Wiley and Sons).

\end{thebibliography}
%\begin{thebibliography}{10}
%\bibitem{Harlow} D. Harlow, S. Shenker, D. Stanford, L. Susskind, ÒEternal Symmetree,Ó [arXiv:1110.0496 [hep-th]].
%\bibitem{Nicolas} Julou T, Mora T, Guillon L, Croquette V, Schalk IJ, Bensimon D, Desprat N (2013) CellÐcell contacts confine public goods diffusion inside Pseudomonas aeruginosa clonal microcolonies. PNAS 110:12577-12582.
%\end{thebibliography}

\end{document}